\documentclass{ws-ijcga}
\usepackage{verbatim, color}
\newcommand{\SA}{\mathcal{S}}
\newcommand{\mb}[1]{\ensuremath{\mathbf{#1}}}

\begin{document}

\markboth{Hummel, Simmerling, Coutsias}
{Laguerre-Intersection Method for Implicit Solvation}

\catchline


\title{LAGUERRE-INTERSECTION METHOD FOR IMPLICIT SOLVATION}

\author{MICHELLE HATCH HUMMEL\footnote{
Currently at Sandia National Laboratories, Albuquerque, NM  87123 USA}}

\address{Department of Applied Mathematics and Statistics, Stony Brook University\\
Stony Brook, NY 11794, USA\,\\
mhhumme@sandia.gov
}

\author{CARLOS SIMMERLING}

\address{Department of Chemistry and Laufer Center, Stony Brook University\\
Stony Brook, NY 11794, USA\, \\
carlos.simmerling@stonybrook.edu
}

\author{EVANGELOS A. COUTSIAS \footnote{
Corresponding author.}}

\address{Department of Applied Mathematics and Statistics and Laufer Center, Stony Brook University\\
Stony Brook, NY 11794, USA\,\\
evangelos.coutsias@stonybrook.edu
}

\author{BIHUA YU}

\address{Department of Applied Mathematics and Statistics, Stony Brook University\\
Stony Brook, NY 11794, USA\,\\
bihua.yu@stonybrook.edu 
}
\maketitle

\pub{Received May 02, 2016}{Revised (revised date)}
{Communicated by Michelle Hatch Hummel}

\begin{abstract}

Laguerre tessellations of macromolecules capture properties such as molecular
interface surfaces, volumes and cavities. Explicit solvent molecular
dynamics simulations of a macromolecule are slow as the number of solvent atoms
considered typically increases by order of magnitude. Implicit methods model the solvent
via continuous corrections to the force field
based on estimates of the solvent exposed surface areas of individual atoms, 
gaining speed at the expense of accuracy. 
However, Laguerre cells of exterior atoms tend to be overly large or unbounded.
Our method, the Laguerre-Intersection method, caps cells in a physically accurate manner by
considering the intersection of the space-filling diagram with the Laguerre tessellation.
This method optimizes an adjustable parameter, the weight, to ensure the areas and volumes
of capped cells exposed to solvent are as close as possible, on average, to those computed
from equilibrated explicit solvent simulations. The contact planes are radical planes,
meaning that as the solvent weight is varied, cells remain constant.  
We test the consistency of our model using a high-quality trajectory of HIV-protease, a dimer
with flexible loops and open-close transitions.  We also compare our results with  
interval-arithmetic Gauss-Bonnet based method.  
Optimal solvent parameters quickly converge, which we use to illustrate the increased accuracy of the
Laguerre-Intersection method over two recently proposed methods as compared to the explicit model.

\end{abstract}

\section{Introduction}

Laguerre tessellation (also known as the weighted Voronoi or power tessellation) is a well known mathematical tool used to partition the space containing
 data points into cells.  
Each data point is assigned the region which 
is closest to the given data point (generator of the cell) by the ``power distance''.  Laguerre methods have been widely used in
biochemistry to find volumes of atoms, molecules, and
amino acid residues (which we call simply ``residues'') in proteins.  They are useful for cavity location and in studying packing and deformation of
proteins. They may also be used in protein structure prediction to check the quality of predicted structures.  
For example, residue volume may be considered an ``intrinsic''
property of amino acid type and is thus a predictable or checkable quantity \cite{Residue_vols}.
Laguerre surfaces have been used as a quick
and parameter free way to measure accessibility of atoms and residues for quantifying exposure
of a molecule with solvent \cite{Residue_vols}.
Residue contacts, which are important for protein structure and stability, have been studied using
Laguerre surfaces.  Interresidue contact surface areas are useful for protein structure analysis
and prediction, and in studying structure-function relationships. Preferential contacts
between amino acid species and atom-atom contact frequencies have 
also been found\cite{Laguerre_protein_contacts}
\cite{constrained_Voronoi}.
Differences in contact areas between model and reference structures have been used as a scoring function
and for benchmarking protein structure prediction methods \cite{CAD_score}. Contact surfaces have also been used to
characterize interactions between multiple proteins, or proteins and ligands, with important applications
in protein docking and formation of complexes
 \cite{dynamic_interface_between_proteins} \cite{interface}. 

Surface area calculations play an increasingly important role in molecular dynamics
simulations of macromolecules. For example, explicit solvent molecular dynamics simulations
of a macromolecule are slow.
The solute is imbedded in an adequately large periodic box filled with solvent and
the system is simulated for a sufficiently long time so solvent molecules achieve
thermal equilibrium.  This typically increases the number of atoms in the system
by an order of magnitude  and furthermore the solvent viscosity significantly 
slows global dynamics such as protein folding \cite{Zagrovic:2003}. Using implicit solvent
in molecular dynamics is an attractive alternative due to the reduced number of atoms in the
simulation, along with the reduced viscosity due to the instantaneous relaxation of solvent
following solute conformational changes. Implicit solvent calculations also play an important
role in the post-processing of MD trajectories that were carried out in explicit water as
a means to estimate binding affinities (the MM-PBSA method \cite{Kollman:2000}).
Here implicit solvent is often advantageous since it directly provides an estimate of the free energy of
(de)solvation, as compared to the enthalpy that is obtained using explicit solvent. This can be valuable in
some cases where solvation free energy is needed, since obtaining this value in explicit solvent requires
proper averaging over all possible explicit solvent configurations for a given solute configuration,
obtaining the entropic contribution as well as the appropriately averaged enthalpy.
Popular implicit solvent models include the Poisson-Boltzmann and Generalized Born methods,
however these model only the polar contribution to solvation energy.
A more accurate treatment involves also including the nonpolar (hydrophobic and dispersion)
contributions to solvation, which are typically considered as approximately proportional
to the solvent accessible surface area of the solute. Several methods have been developed
and are commonly used \cite{Connolly:1983} \cite{Tan:2007} \cite{Weiser:1999}.
The speed and accuracy of these methods is vital to successful application,
but due to the computational overhead many recent studies of protein folding
omit nonpolar solvation despite the importance of the hydrophobic effect \cite{Nguyen:2014}.

The Laguerre cell is the region which
is closest to the generator of the cell (atom or residue center, for example) by the power distance.
The power distance is an appropriate quantity to use as larger areas are assigned to data points with larger
weights (larger atoms or residues, for example). The Laguerre cell of a data point depends on its neighbors.
This poses a problem
for systems that are typically found in a medium such as solvent but for which the surrounding
medium is not explicitly known. In this situation, Laguerre cells of points on the convex hull
are unbounded and the cells of points otherwise in contact with the surrounding medium tend
to be too large. To arrive at a geometrical description of a macromolecule that is
both independent of the rapidly changing detailed structure of the surrounding solvent
molecules (typically water) while at the same time encoding an "average" geometrical
profile as seen by the medium is a challenge that must be overcome in order to enhance the accuracy of implicit solvation modeling.

Researchers have addressed this difficulty in a variety of ways. Some
consider only cells in the bulk of the protein which are not affected by the surrounding (unknown)
environment. Others
surround the structure by a layer of water or an artificial environment of spheres with size equal to the average amino
acid size \cite{Volume_atoms_protein_surf_Voronoi} \cite{Lag_decomp_app_protein_folds}.
However, this causes the size of the problem to increase, typically by an order of magnitude. Soyer et al., bounded Laguerre cells by only
considering Laguerre facet vertices from tetrahedra of a small enough
size \cite{surf_vs_bulk_Voronoi}. Still cells near the boundary tend to be
elongated or have fewer facets than those in the bulk of the protein.  Cazals method 'Intervor'
constructs interface surfaces by considering facets of dual Delaunay edges in the 
space filling diagram and discarding any edges that belong to large tetrahedra \cite{Cazals:2006} \cite{Cazals:2010}.
Discarding these edges causes
pertinent information to be lost.

Another class of models prunes and truncates certain Laguerre facets.
Mahdavi et al. construct intermolecular interface surfaces by considering truncated facets 
in the union of extended convex hulls \cite{Mahdavi:2012} \cite{Mahdavi:2013}.  The convex hulls
are extended in a fashion to mimic a 1.4 Angstrom solvent radius.
Ban and coworkers construct a series of interface surfaces based on a sequence of alpha complexes
and truncate facets to within alpha complex tetrahedra, but no attempt is made to 
mimic the effect of the solvent \cite{Ban:2006} \cite{Headd:2007}.

McConkey et al.
construct Laguerre-like cells by considering the surfaces of extended radical contact
planes between neighboring atoms within a cutoff distance and the surface of an
expanded sphere \cite{constrained_Voronoi}. However, the algorithm miscalculates Laguerre volumes when
the atom center lies outside its cell (what we term an ``engulfing contact''),
a situation that our algorithm treats correctly (See Section \ref{signed_volume}).  Cazals et al. in their paper
``Computing the volume of a union of balls: a certified algorithm'' \cite{Cazals:2011}
provide a formalization of the McConkey's algorithm which also correctly calculates engulfing
contact volumes. While these algorithms compute the 
restriction of Laguerre cells to the space-filling model, there are three key differences with our model.  McConkey's and Cazals' 
algorithms are based on the Gauss-Bonnet theorem whereas our algorithm is an inclusion-exclusion model which is an extension 
of Edelsbrunner's algorithm for calculating atomic and molecular surface areas and volumes \cite{Union}.  Inclusion-exclusion contributions extended to per-restriction quantities 
are given in equations (12)-(28) and (34)-(36). The inclusion-exclusion method
is advantageous in that cycles of intersection points on the surface of the sphere do not need to be calculated which decreases algorithm complexity.
The second key difference is that the former algorithms expand nonsolvent atoms 
by a solvent radius of $1.4$  \AA\
while our algorithm expands the squared radius. 
Expanding cells' radii means that cells are bounded by extended radical planes rather
than radical planes. This
poses problems since the cells of atoms in the bulk, which should not be affected by the cutoff distance,
are affected. Furthermore, cells of small atoms completely disappear 
for cutoff distances as small as 1.4 \AA.  The third difference is that Cazals' resorts to interval arithmetic and/or exact arithmetic which 
increases runtime by a factor of ten over the floating point models.  As we are interested in calculating Laguerre-Intersection quantities at each step in a molecular dynamics (MD)
simulation, speed is of utmost importance, and we can achieve desired accuracy using floating point operations.

The proposed Laguerre-Intersection cell algorithm \cite{Hummel:2014}, considers the intersection of
Laguerre cells with expanded atoms. The
contact planes are the radical planes, which means that as the solvent weight is varied, Laguerre
cells stay constant. This method simulates the environment better than using the extended radical plane.
The Laguerre cells are capped in a physical manner that enables the study of boundary facets, rather than 
truncation by convex hulls.

We calculate molecule specific optimal solvent parameters from explicit solvent HIV protease
trajectory data.  The HIVP
dimer has two flexible loops covering the active site that 
exhibit open-closed dynamics \cite{Hornak:2005}.  We generated a 100 ns trajectory of HIV 
in explicit water to use as reference data (see Results).  Despite the configurational 
variability, optimal solvent parameters converge quickly which we use to predict 
Laguerre quantities in the remainder of the trajectory.  We show that these predicted quantities 
are closer to those found using explicit models than two current alternative methods.   Further work is required to determine optimal solvent 
parameters for an arbitrary molecule.


\section{Background}
\subsection{Laguerre and Laguerre-Intersection cells}
\begin{figure}
\begin{center}
\includegraphics[scale=.2]{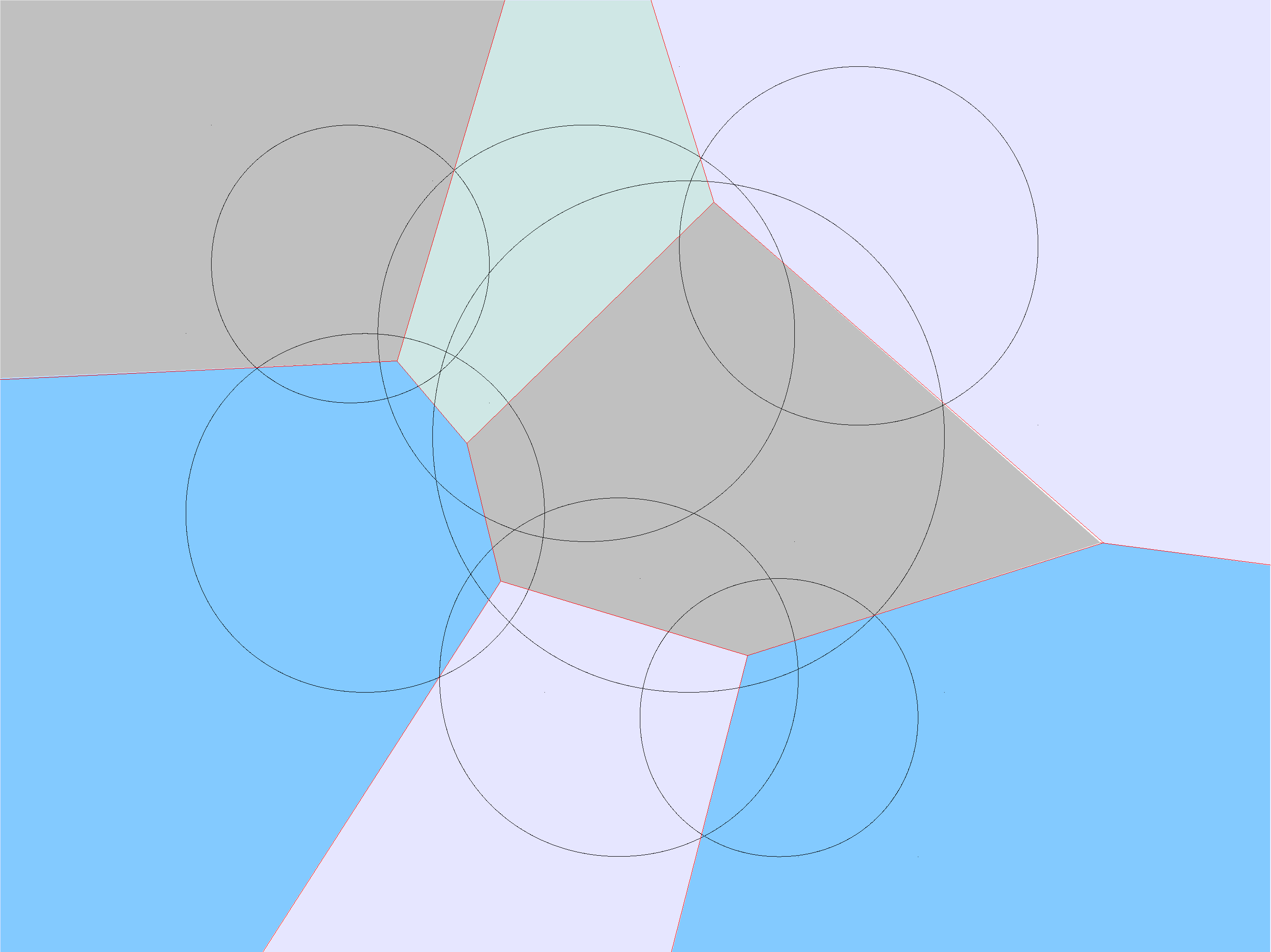}
\includegraphics[scale=.35]{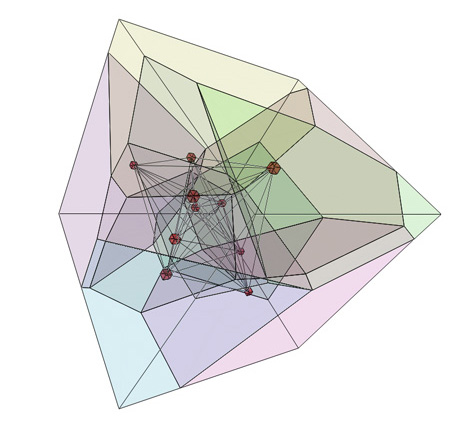}
\end{center}
\caption{2d and 3d Laguerre diagram: 2d illustrates how the radical plane partitions space between
 atoms of different radii; 3d courtesy of Frederick Vanhoutte. \label{laguerre_diagram}}

\end{figure}

Consider a molecule represented by a set of spheres or atoms
$\mathcal{A} \subset \mathbb{R}^3 \times \mathbb{R}$. For $p_i \in \mathcal{A}$ we write
$p_i=(p_i', w_i)$ where $p_i'$ is the center of the atom and $w_i=r_i^2=p_i''$ is the weight or squared
radius of the atom.
The Laguerre cell, $L_i$, of atom $i$ is defined to be
\begin{equation}
L_i=\{ x' \in \mathbb{R}^3 \quad :\quad |p'_i-x'|^2-w_i \quad \leq \quad|p'_j-x'|^2-w_j
\quad \text{ for all } \quad p_j \in \mathcal{A} \}
\end{equation}
and is a convex polytope
(Fig. \ref{laguerre_diagram}).

The quantity $|p_i'-x'|^2-w_i$ is called the power distance 
between
$p_i$ and $x'$ and is written
as
\begin{equation}
\pi(p_i,x')=|p_i'-x'|^2-w_i.
\end{equation}
A Laguerre cell is the set of points whose nearest neighbor by the power distance
is $p_i$, and each Laguerre facet lies on the plane which is equi-powerdistant between two points and which
is called the radical plane.

\begin{figure}
\begin{center}
\includegraphics[scale=.25]{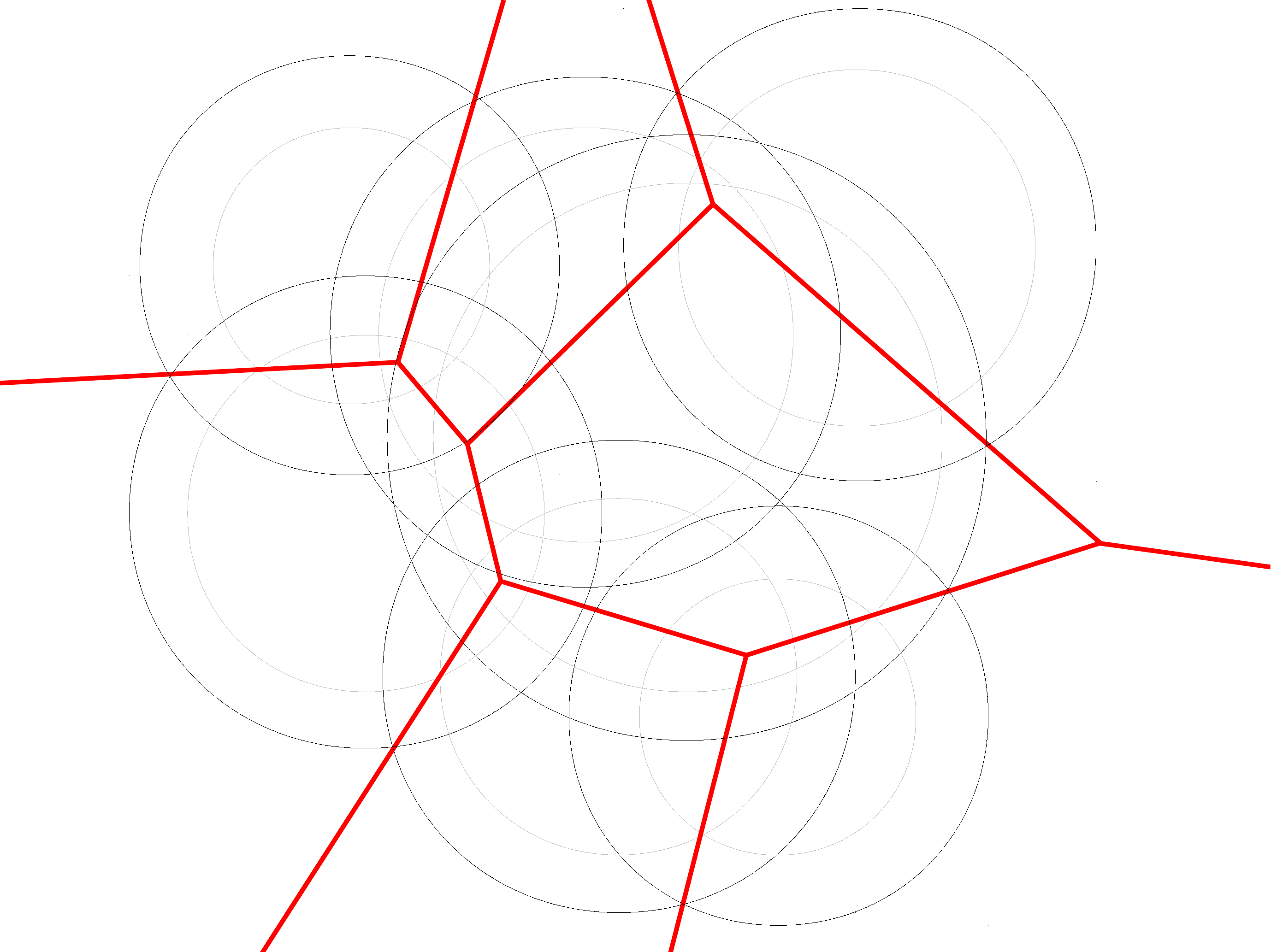}

\end{center}
\caption{Red bounds Laguerre cells of $\mathcal{A}(w)$ (shown in black)
and $\mathcal{A}(0)$ (shown in gray). Note that the cells are the same for both weights. 
This is the reason that varying the weight rather than the radius in the space filling diagram gives 
better agreement between Laguerre-Intersection quantities and Laguerre quantities found using explicit solvent (See
Section \ref{radius_vs_weight}).
}
\label{Laguerre_vary_weight}
\end{figure}

The weights of the atoms in $\mathcal{A}$ may be increased or decreased by a certain amount $w$ which
we call the solvent weight. In this section, we call this modified set
$\mathcal{A}(w)$ where $p_i(w) \in \mathcal{A}(w)$ has $p_i'(w)= p_i'$ and $w_i(w)=w_i+w$.
The Laguerre cell of $p_i(w)$ is defined as
\begin{equation}
L_i(w)=\{ x' \in \mathbb{R}^3 : |p_i(w)-x'|^2-w_i(w)
\leq |p_j(w)-x'|^2-w_j(w) \quad \forall p_j(w) \in \mathcal{A}(w) \}.
\end{equation}
Since $L_i(w)=L_i(0)$ for all $w$ (See Fig. \ref{Laguerre_vary_weight}) we simply write the Laguerre cell of atom $i$ as $L_i$.

The Laguerre diagram, $\mathcal{L}(\mathcal{A})$, of $\mathcal{A}$ is the collection of all Laguerre
cells and their faces which we call Laguerre facets, segments, and nodes. A Laguerre facet is the intersection
of two Laguerre cells and is a subset of the plane which is equi-powerdistant from the generator of the
two cells. A Laguerre segment is the intersection of at least three Laguerre cells, and a Laguerre node
is the intersection of at least four Laguerre cells.
For $p_i \in \mathcal{A}$ define
\begin{equation}
B_i(w)= \{ x \in \mathbb{R}^3 : |p_i'-x|^2-w_i(w) \leq 0 \}.
\end{equation}
The space filling model of $\mathcal{A}$ with solvent weight $w$ is defined as
\begin{equation}
\mathcal{B}(w)=\bigcup B_i(w)
\end{equation}
We also define the following quantities:
\begin{itemize}
\item $LV_i$: Volume of the Laguerre cell of atom $i$.
\item $LS_i$: Surface area of the Laguerre facets of atom $i$.
\end{itemize}
A residue is the collection of atoms in a single amino acid unit in the protein.
The Laguerre volume of residue $j$ (residual volume) is the sum of Laguerre volumes of atoms in residue $j$.
The interresidue Laguerre surface area between residues $j$ and $k$ is the sum of
areas of Laguerre
facets which lie between atoms in residue $j$ and atoms in residue $k$.

\subsection{Laguerre diagram and regular tetrahedrization}
The regular tetrahedrization is dual to the Laguerre diagram $\mathcal{L}(\mathcal{A})$.
There is a one to one correspondence between the $(3-k)$-faces in $\mathcal{L}(\mathcal{A})$ and the $k$-simplices
in $\mathcal{T}(\mathcal{A})$.
Each node in $\mathcal{T}(\mathcal{A})$ corresponds to a tetrahedron
in $\mathcal{T}(\mathcal{A})$ whose vertices are equipowerdistant to the node,
which we call the characteristic
point of the tetrahedron.
Each segment in
$\mathcal{L}(\mathcal{A})$ corresponds to a triangle in $\mathcal{T}(\mathcal{A})$ whose vertices are equipowerdistant
to the power segment. For each facet in $\mathcal{L}(\mathcal{A})$ there is an edge
in $\mathcal{T}(\mathcal{A})$ whose vertices are equipowerdistant to the power facet.
Each cell in $\mathcal{L}(\mathcal{A})$
corresponds to a point in $\mathcal{T}(\mathcal{A})$, namely the generator of the cell.
We call a point whose center is on the convex hull of $\mathcal{T}(\mathcal{A})$ exterior to the tetrahedrization.
Otherwise a point is called interior to $\mathcal{T}(\mathcal{A})$.
An edge is called exterior if both of its vertices are exterior to $\mathcal{T}(\mathcal{A})$.
An edge is called interior if at least one of its vertices is interior.
Note that the Laguerre cell of an exterior point is unbounded whereas the Laguerre cells of interior
points are bounded. The Laguerre facets corresponding to exterior edges are unbounded whereas Laguerre
facets of interior edges are bounded.

In this paper we take for granted robust algorithms for the calculation of the 
regular tetrahedrization 
and the $\alpha$-complex which is a subset of the the tetrahedrization.

\section{Methods}
\subsection{Computation of Laguerre surface areas of interior facets}\label{Interior_cells}
\begin{figure}
\begin{center}
\includegraphics[scale=.31]{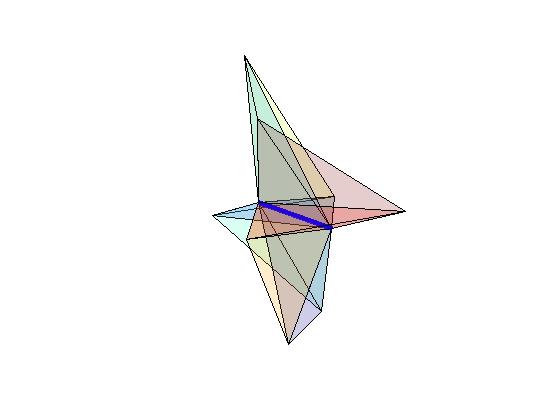} \includegraphics[scale=.31]{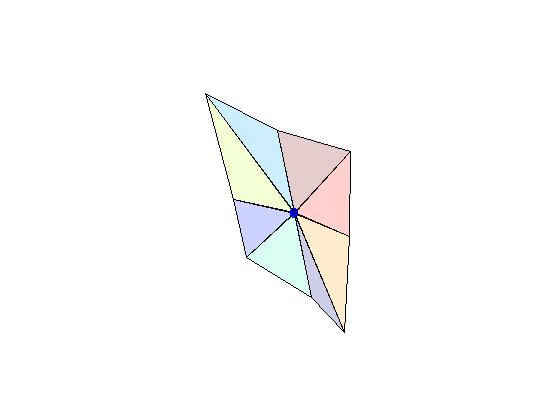}
\end{center}
\begin{center}
\includegraphics[scale=.28]{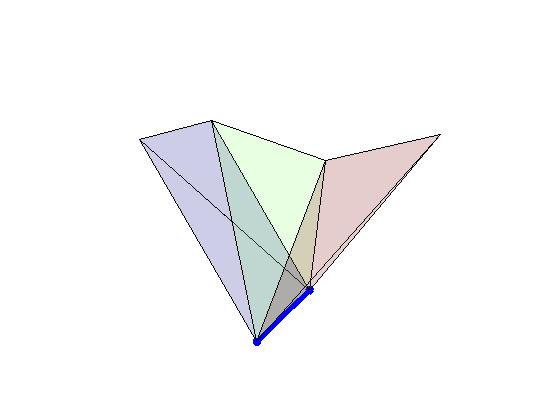} \includegraphics[scale=.28]{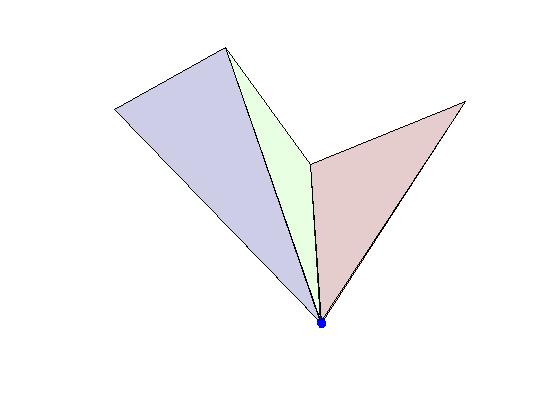}
\end{center}
\caption{Top row: Side and perpendicular views of the complete tetraring of an
interior edge (shown in dark blue). Bottom row: Side and perpendicular views of an incomplete tetraring
of an exterior edge (shown in dark blue).}
\label{tetra_ring}
\end{figure}

Interior Laguerre cells are bounded by Laguerre facets. Each of these facets correspond to an edge
in the regular tetrahedrization. We call an edge $e_{ij}$ if the vertices of that edge are $p'_i$ and $p'_j$ and
call the Laguerre facet corresponding to that edge $L_{ij}$. The facets are convex
polygons whose vertices are nodes in the Laguerre diagram, namely the characteristic points of the
tetrahedra which surround that given edge. If a Laguerre cell has $n$ facets, the volume may be
divided into $n$ pieces, one for each of its facets. The volume of the Laguerre cell of $i$ corresponding to the edge $e_{ij}$ is written $LV^{(i)}_{ij}$
and the volume of the Laguerre cell of $j$ corresponding to the edge is written $LV^{(j)}_{ij}$.
We call the ordered list of tetrahedra which surround
a given edge, i.e. the ordered list of tetrahedra in the edge's star,
a \emph{tetrahedra ring} or \emph{tetraring}. A tetraring is called complete
if the ring closes, otherwise it is called incomplete. Note that the tetraring
of an interior edge is complete, whereas the tetraring of an exterior edge is incomplete
(Fig. \ref{tetra_ring}).
In this section, we assume that the protein is surrounded by a layer of solvent. This means that
all nonsolvent atoms are interior to the regular tetrahedrization as well as all
edges which contain a nonsolvent atom as a vertex. We call these ``nonsolvent edges''.
The computation of the Laguerre
volumes and surfaces of each nonsolvent atom proceeds as follows:
\begin{itemize}
\item Regular tetrahedrization of all atoms is computed
\item Characteristic points of tetrahedra are found
\item For each nonsolvent edge
\begin{enumerate}
\item Surface area of the corresponding Laguerre facet is computed and assigned to appropriate atoms
\item Using the calculated area, corresponding Laguerre volumes are found and assigned to appropriate atoms
\item Interresidual Laguerre surface areas are assigned
\end{enumerate}
\item Laguerre residue volumes are summed
\end{itemize}

\subsubsection{Surface areas}
\begin{figure}
\begin{center}
\includegraphics[scale=.3]{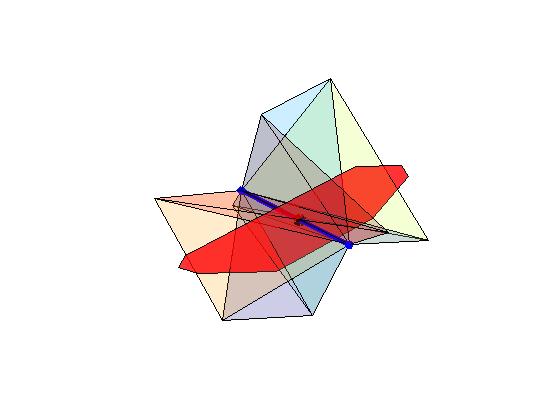}
\includegraphics[scale=.3]{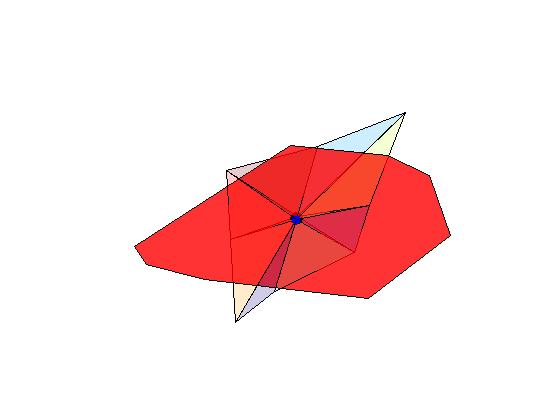}
\end{center}
\begin{center}
\includegraphics[scale=.3]{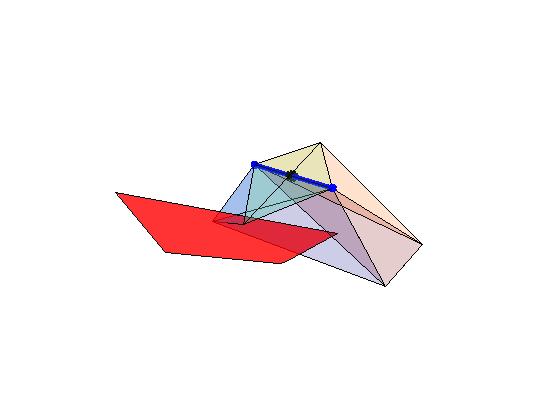}
\includegraphics[scale=.3]{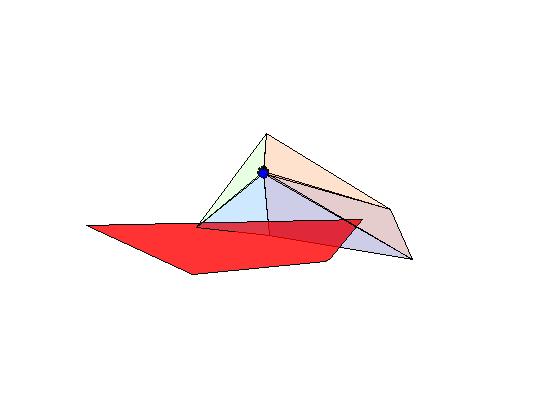}
\end{center}
\caption{Laguerre facet of edge represented in red. The vertices of the facet are the characteristic
points of the tetrahedra in the edge's tetraring.
Top row: Two views of a facet which is intersected by its corresponding edge.
Bottom row: Two views of a facet which is not intersected by its corresponding edge.}
\label{edge_facet}
\end{figure}

\begin{figure}
\begin{center}
\includegraphics[scale=.3]{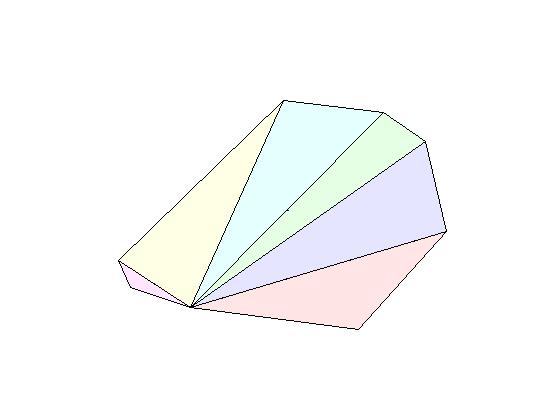}
\includegraphics[scale=.3]{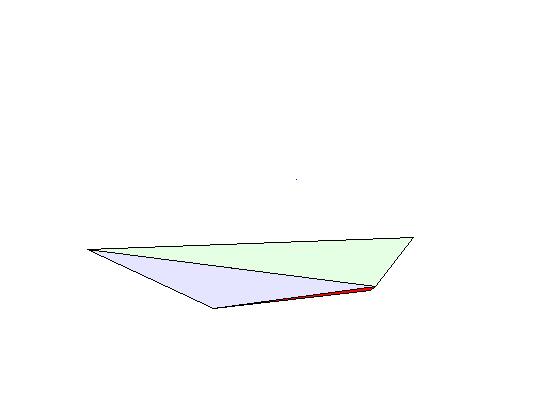}
\end{center}
\caption{Subdivision method for the two facets shown in Fig. \ref{edge_facet}.}
\label{subdivide_facet_1}
\end{figure}

When considering
edge $e_{ij}$, we compute the surface area, $LS_{ij}$, of the facet, $L_{ij}$, between atoms $i$ and $j$.
The ordered vertices in the Laguerre facet are the characteristic points of the tetrahedra
in the edge's tetraring. Note that the edge does not always intersect its Laguerre facet, but
it is always perpendicular to it (Fig. \ref{edge_facet}).
Next we define the characteristic point of an edge. The characteristic
point, $x_{ij}=(x_{ij}',x_{ij}'')$ of an edge, $e_{ij}$ satisfies
\begin{equation}
\pi(p_i,x_{ij}')=\pi(p_j,x_{ij}')
\end{equation}
with $x_{ij}''$ minimal. The center of this point lies on the intersection of the line containing $p_i'$ and $p_j'$
and the plane of points which
are equipowerdistant to $p_i$ and $p_j$.
The method computes the surface area as the sum of the areas of the triangles
as shown in Fig. \ref{subdivide_facet_1}.

\subsubsection{Volumes}
\begin{figure}
\begin{center}
\includegraphics[scale=.3]{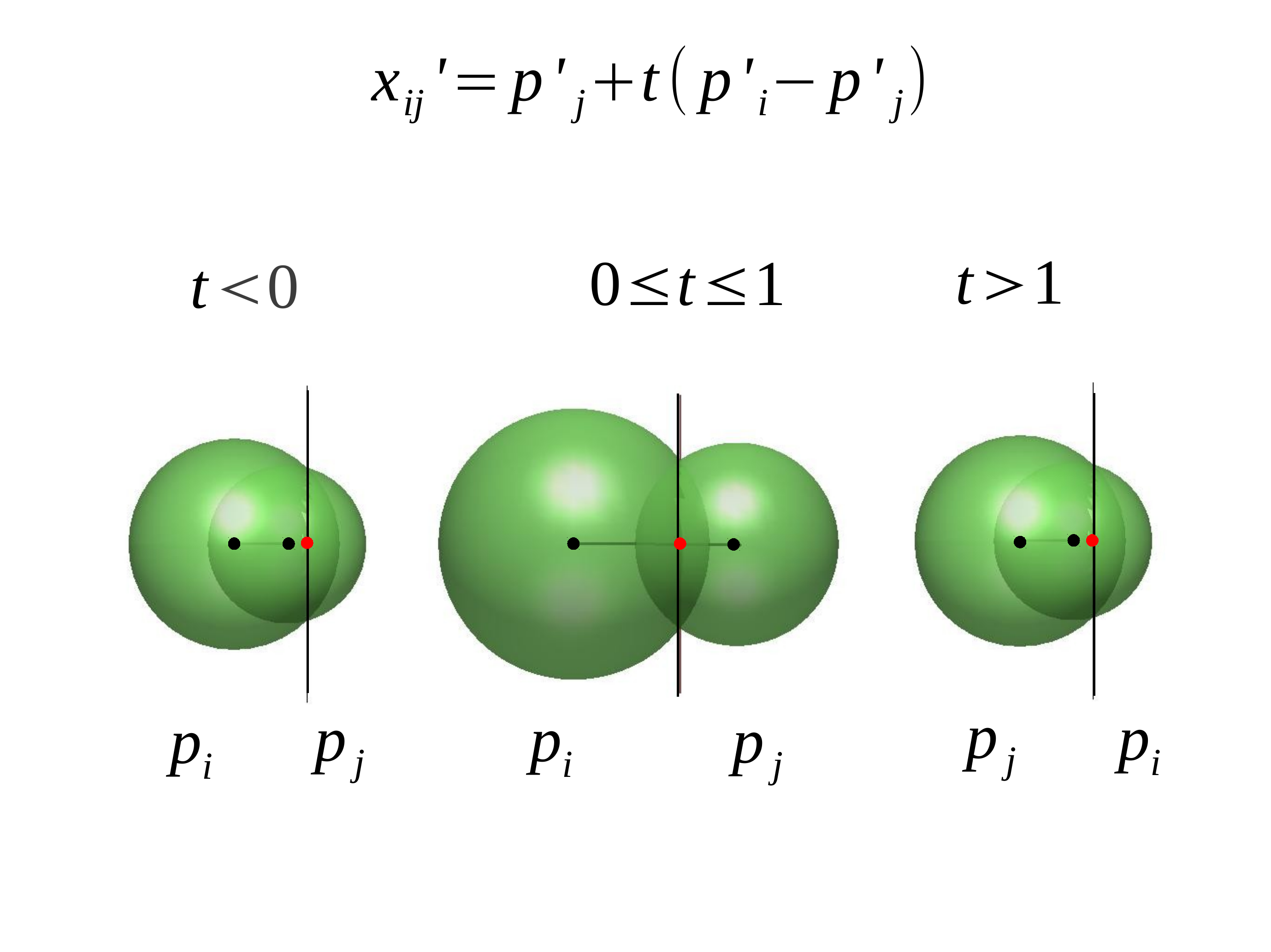}
\end{center}
\caption{The value of $t$ is determined by the relationship of the equi-powerdistant plane with
respect to the two edge points and indicates if the center of the generator of the cell is
interior or exterior to its cell.}
\label{t_val}
\end{figure}

\begin{figure}
\begin{center}
\includegraphics[scale=.3]{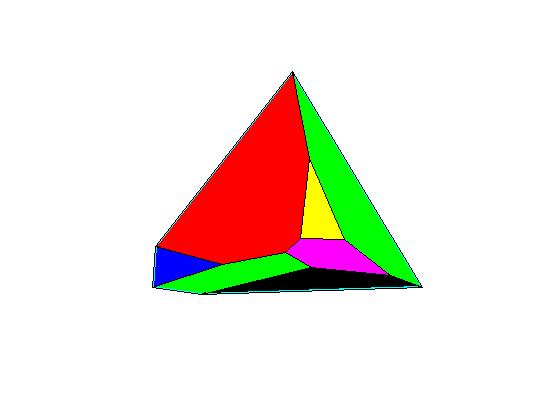}
\includegraphics[scale=.3]{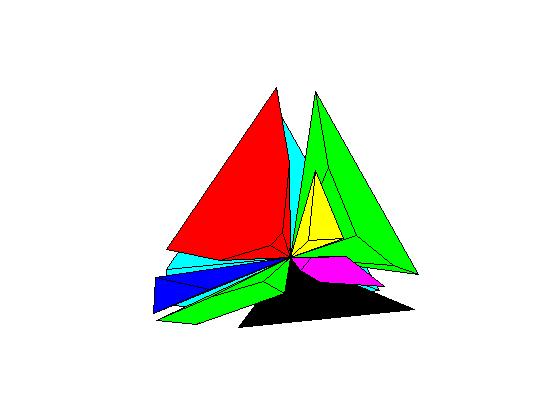}
\end{center}
\caption{Left: Laguerre cell that contains generator's center. Right: Pyramidal decomposition of Laguerre cell. All volumes are positive.}
\label{pyramidal_decomp_1}
\end{figure}

\begin{figure}
\begin{center}
\includegraphics[scale=.21]{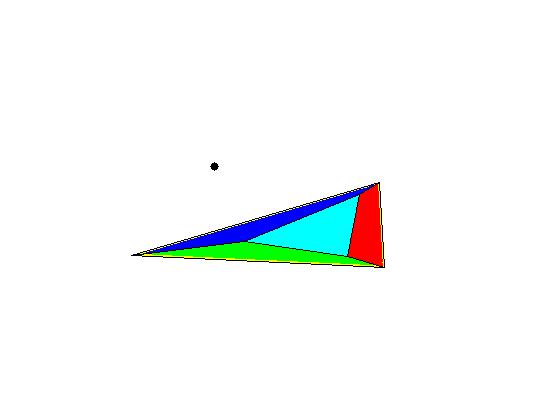}
\includegraphics[scale=.21]{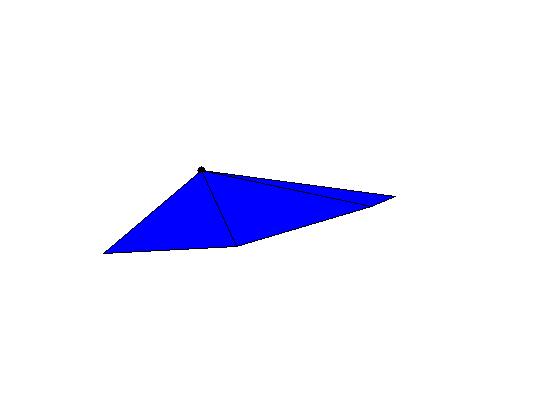}
\includegraphics[scale=.21]{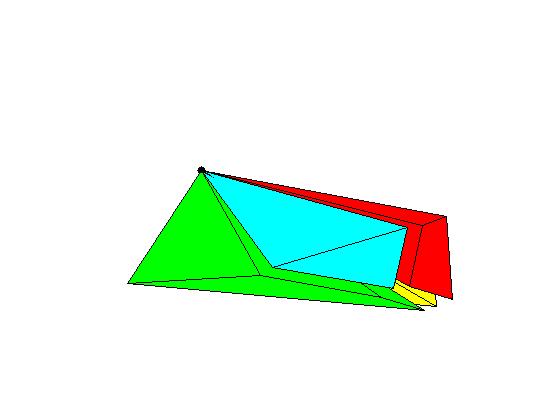}
\end{center}
\caption{Left: Laguerre cell that does not contain generator's center (black vertex).
Center: Negative pyramidal volume. Right: Positive pyramidal volumes.}
\label{pyramidal_decomp_2}
\end{figure}

The Laguerre volume contributions, $LV^{(i)}_{ij}$ and $LV^{(j)}_{ij}$,
are pyramids with base $L_{ij}$, and heights
$h^i_{ij}$ and $h^j_{ij}$ respectively, with
\begin{eqnarray}
h^i_{ij} & = & |x_{ij}'-p_i'| \\
h^j_{ij} & = & |x_{ij}'-p_j'|.
\end{eqnarray}
We can write
\begin{equation}
x_{ij}'= p_j'+t(p_i'-p_j').
\end{equation}
Thus the position of the plane containing the edge's facet with respect to $p_i'$ and $p_j'$
is encoded in the value of $t$ (Fig. \ref{t_val}).
This gives
\begin{eqnarray}\label{signed_volume_equation}
LV^{(i)}_{ij}&=&sign(1-t)\frac{h^i_{ij}LS_{ij}}{3}  \\
LV^{(j)}_{ij}&=&sign(t)\frac{h^j_{ij}LS_{ij}}{3} \nonumber
\end{eqnarray}
The term $LV^{(i)}_{ij}$ is positive if the corresponding pyramid lies in $L_i$ and
negative otherwise (Figs. \ref{pyramidal_decomp_1}, \ref{pyramidal_decomp_2}).  We call the 
latter case an ``engulfing'' contact; that is the generator of one cell lies in the Laguerre 
cell of another atom.

Formula \ref{signed_volume_equation} is in contrast to the volume calculation
(Equation 7b) in McConkey's method \cite{constrained_Voronoi} which does not consider signed volumes
and thus returns incorrect quantities for engulfing contacts (For example, see Fig. \ref{pyramidal_decomp_2}).
\subsection{Removing solvent}\label{signed_volume}
\begin{figure}
\begin{center}
\includegraphics[scale=.3]{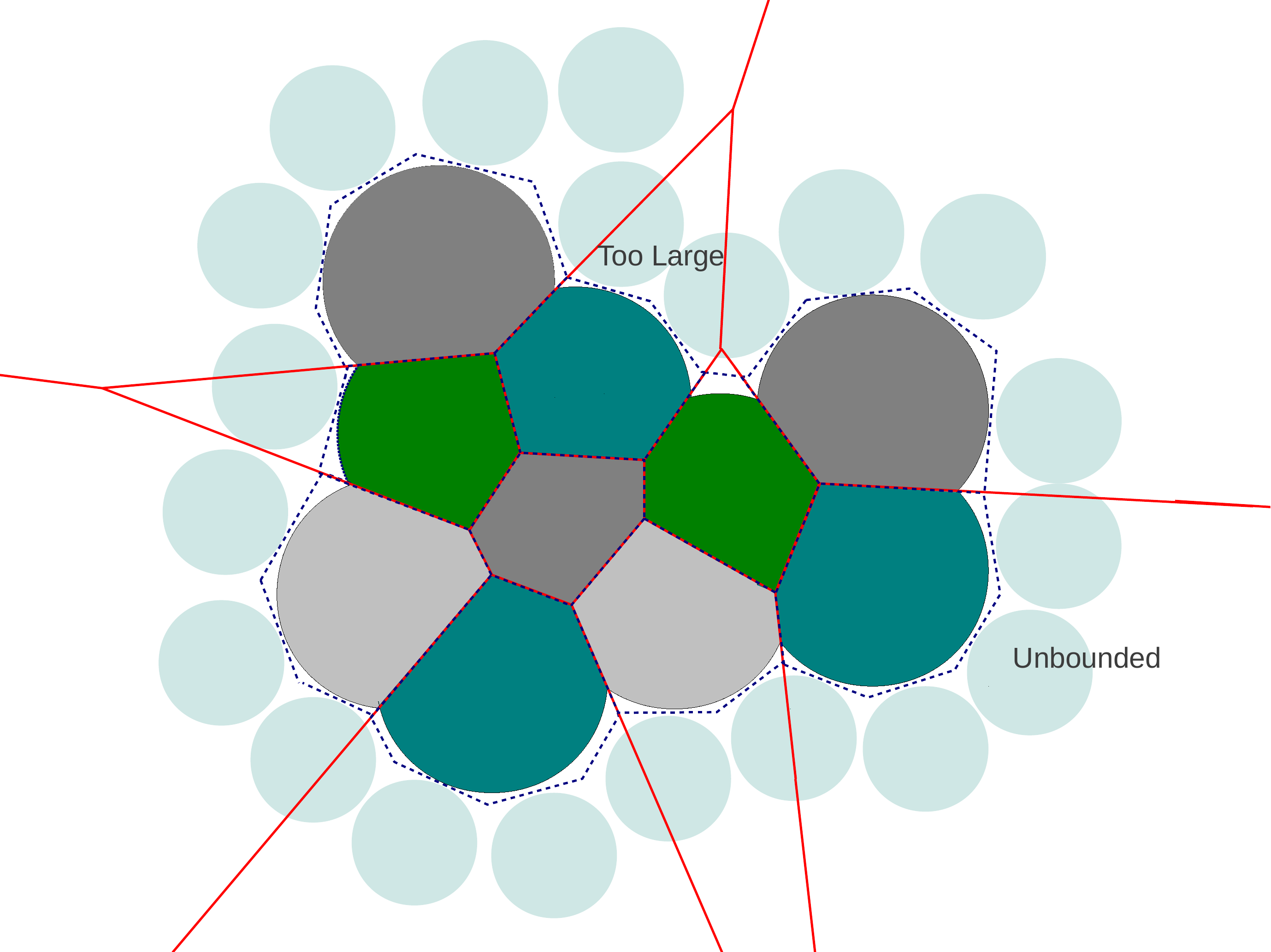}
\end{center}
\caption{Example of Laguerre-Intersection cells (shown in grays and greens). Solvent is
represented by light blue spheres, boundary of Laguerre cells of molecule
(minus solvent) are shown in red, and boundary of Laguerre cells of molecule in solvent is
represented by the dotted line.}
\label{solvent_cartoon}
\end{figure}

\begin{figure}
\begin{center}
\includegraphics[scale=.35]{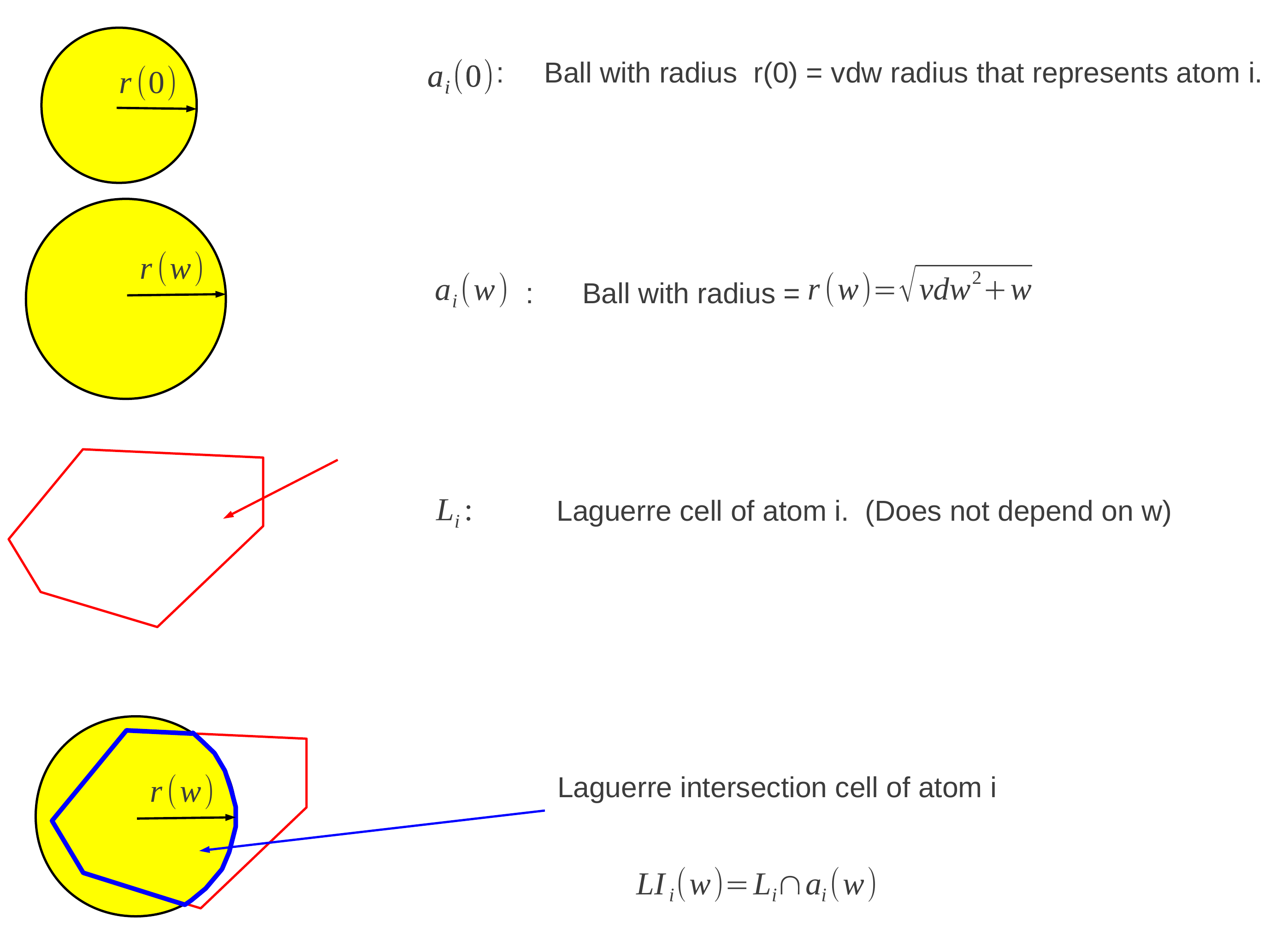}
\end{center}
\caption{Laguerre cell does not depend on $w$ while the Laguerre-Intersection cell does depend on $w$.}
\label{independent_w}
\end{figure}

We would like to compute Laguerre-like volumes and surfaces of a molecule that is not surrounded
by solvent, so that these quantities are as close as possible to the Laguerre volumes and surfaces that
are found when the molecule is in its typical environment. In doing this, we come across two problems:
\begin{itemize}
\item The Laguerre cells are unbounded for certain atoms on the convex hull of $\mathcal{A}'$.
\item The Laguerre cells of atoms that would otherwise be in contact with solvent are too large
\end{itemize}
This may be overcome by using Laguerre intersection
cells, $LI(w)$. The Laguerre intersection cell of an atom is the intersection of the Laguerre
cell of that atom with the expanded atom.
\begin{equation}\label{LI}
LI_i(w)=L_i \bigcap B_i(w).
\end{equation}
Exterior cells are bounded in a realistic way, and the cells of
atoms which would otherwise be in contact with solvent are shrunk to a more
appropriate size (Fig. \ref{solvent_cartoon}).
Laguerre cells are constant as a function of $w$, which means this parameter can be tuned
to generate appropriate exterior Laguerre intersection cells while interior
cells remain unchanged (Fig. \ref{independent_w}).

Define $LIS_i(w)$ and $LIV_i(w)$, be the surface area and volume of $LI_i(w)$
\begin{eqnarray}
 LIS_i(w) & = & surf(LI_i(w)) \nonumber \\
 LIV_i(w) & = & vol(LI_i(w)). 
\end{eqnarray}
Just as the
individual atomic surface area and molecular volume can be written as a sum of contributions
corresponding to simplices in the alpha complex using the inclusion-exclusion formulas
(See \ref{inclusion_exclusion})
, $LIS$ and $LIV$ can
be split into contributions (Eqs. \ref{ie_LIV}, \ref{ie_LIS}) from simplices in the alpha complex which is a subset
of the Delaunay tetrahedrization  \cite{WAlpha}.
Define $\mathcal{C}(w)$ to be the alpha complex
of $\mathcal{A}(w)$ with $\partial\mathcal{C}(w)$ the set of simplices in the boundary of
$\mathcal{C}(w)$.
For $T \subset \mathcal{A}(w)$, $\sigma_T$ represents
the simplex which is the convex hull of the centers of points in $T$.
As we are working in three dimensions we only consider $|T|\leq 4$, where $|T|$ is the number of
elements in $T$. Recall that $e_{ij} = \sigma_T$ for $T=\{p_i,p_j\}$. We also represent vertices and triangles
as $v_i$ and $t_{ijk}$ in a similar manner.

The Laguerre-Intersection cell, $LI$, of an atom interior to $\mathcal{C}$ is precisely
the Laguerre cell $L$.    
The Laguerre-Intersection cell of an atom exterior $\mathcal{C}(w)$ consists of 
polyhedral components (contributions from interior simplices) and spherical components 
(contributions from simplices on $\partial \mathcal{C}(w)$).

\begin{figure}[h]
\begin{center}
\includegraphics[scale=.2]{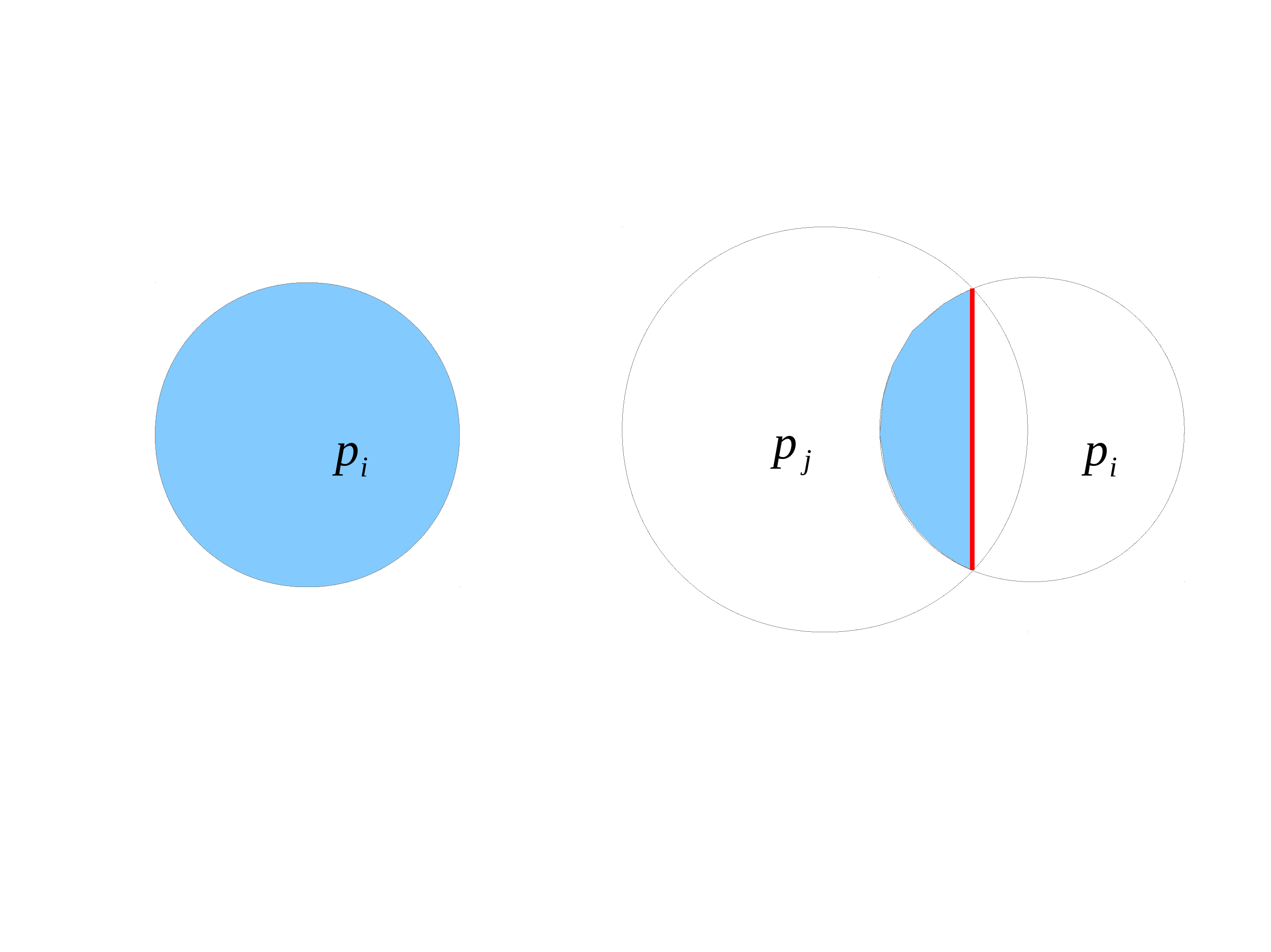}
\includegraphics[scale=.15]{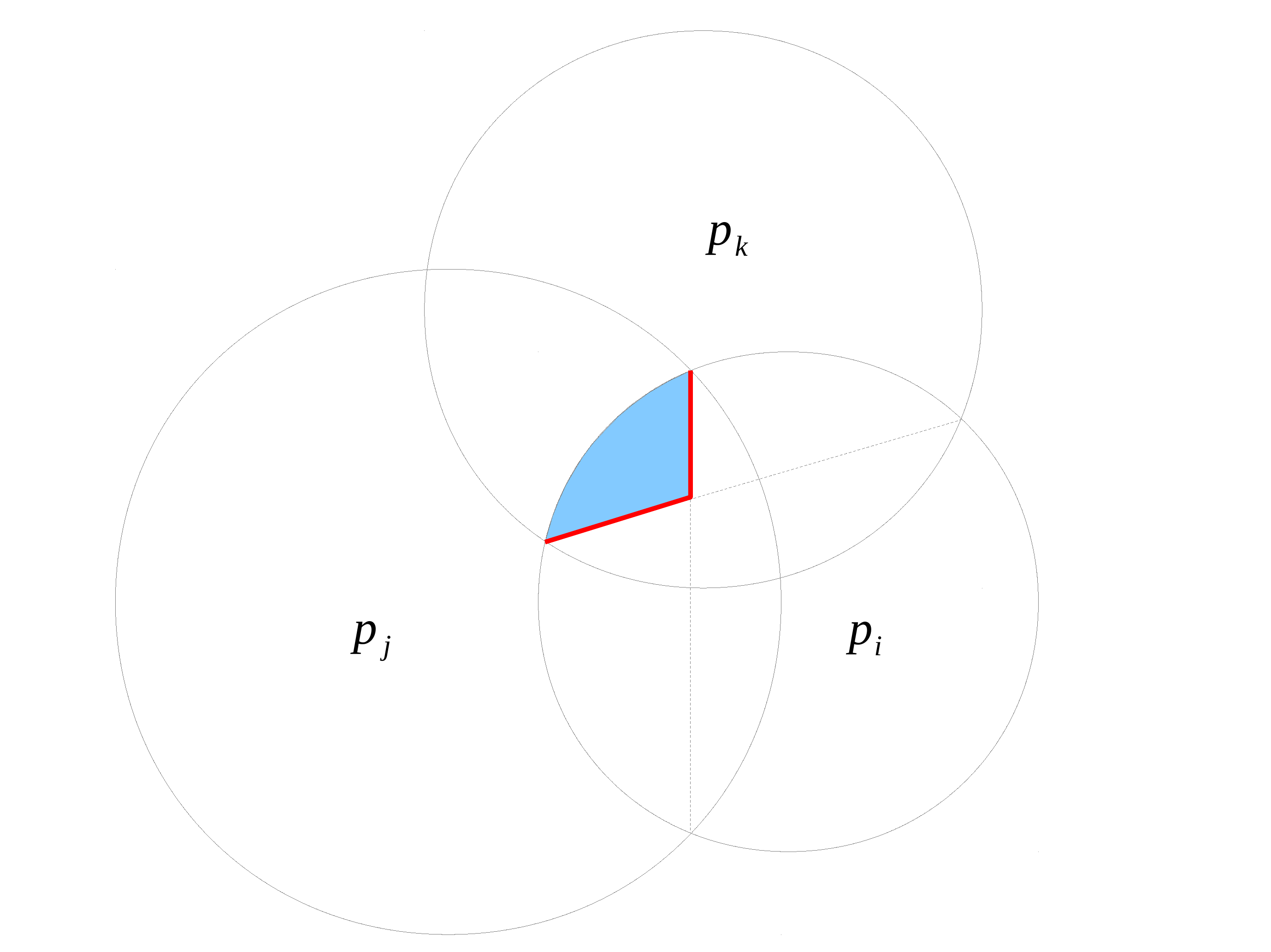}
\end{center}
\caption[Laguerre-Intersection cell volume and surface contributions from vertex, edge,
and triangle on $\partial \mathcal{C}(w)$.]
{Laguerre-Intersection cell volumes (blue), $LIV_i^{(i)}$,
$LIV_{ij}^{(i)}$, $LIV_{ijk}^{(i)}$, and surfaces (red) $LIS_{ij}$, and $LIS_{ijk}$. to
atom $i$ from $\partial \mathcal{C}(w)$ simplices. }\label{LIS_LIV_fig}
\end{figure}

The terms
$LIS_T^{(i)}(w)$ and $LIV_T^{(i)}(w)$ are
the surface area and volume contributions of the simplex $\sigma_{T}$ to $LIV_i(w)$ and $LIS_i(w)$
(the analogs to $S_T^{(i)}$ terms in \ref{surf_atom}).
We can also write $LIV_T^{(i)}(w)=LIV_i^{(i)}(w)$ for $T=\{p_i(w)\}$,
$LIV_T^{(i)}(w)=LIV_{ij}^{(i)}(w)$ for $T=\{p_i(w),p_j(w)\}$,
etc., and likewise for surface areas (See Fig. \ref{LIS_LIV_fig}). The terms $P_T^{(i)}$
and $F_T$ for $|T|=2$ are pyramidal volumes and facet surface areas that will be discussed
shortly.
We have
\begin{eqnarray}\label{ie_LIV}
LIV_i(w)& = &\sum_{\sigma_T \in \partial \mathcal{C}(w)} (-1)^{k+1}c_T LIV_T^{(i)}(w)
\quad \quad |T|=k \nonumber \\
& & + \sum_{\sigma_T \in \mathcal{C}(w)} P_T^{(i)}(w) \quad \quad |T|=2
\end{eqnarray}
and
\begin{eqnarray}\label{ie_LIS}
LIS_i(w) & = &S_i(w)+\sum_{\sigma_T \in \partial \mathcal{C}(w), k>1}
(-1)^{k}c_T LIS_T^{(i)}(w) \quad \quad |T|=k \nonumber \\
& & + \sum_{\sigma_T \in \mathcal{C}(w)} F_T(w) \quad \quad |T|=2
\end{eqnarray}
where $S_i(w)$ is the accessible surface area of atom $i$ with solvent radius $w$, and $c_T$s are given 
for the following simplices
\begin{itemize}
\item $|T|=1$, i.e. a vertex $v_i$:  $c_T=\Omega_T$ is the fraction of the ball $i$ outside the tetrahedra in the alpha complex.  That is
$\Omega_T$ is 
the normalized outer solid angle subtended by the union of tetrahedra in $\mathcal{C}$ which contain $v_i$. 
\item $|T|=2$, i.e. an edge $e_{ij}$: $c_T=\Phi_T$ is normalized outer dihedral angle of 
the union of tetrahedra in $\mathcal{C}$ which contain the edge $e_{ij}$.  

\item $|T|=3$, i.e. a triangle $t_{ijk}$: $c_T$ is 1 if the triangle is singular and $\frac{1}{2}$ if the triangle is regular.  In other words, $c_T$ is the fraction of $V_T$ and $S_T$
that is outside the union of tetrahedra in the alpha complex.
\end{itemize}  
Here $v_i=p'_i$, $e_{ij}=conv(\{p_i',p_j'\})$, and $t_{ijk}=conv(\{p_i',p_j',p_k'\})$.

Since
\begin{equation}
S_i(w)=\sum_{\sigma_T \in \partial \mathcal{C}(w)} (-1)^{k+1} c_T S_T^{(i)}(w), \quad \quad |T|=k
\end{equation}
where $S_T^{(i)}(w)$ is the contribution of $\sigma_T$ to $S_i(w)$,
we can write
\begin{eqnarray}\displaystyle
LIS_i(w) &= & c_i S_i^{(i)}(w) \nonumber \\
& & +\sum_{\sigma_T \in \partial \mathcal{C}(w), k>1}(-1)^k c_T
( LIS_T^{(i)}(w)-S_T^{(i)}(w)) \quad \quad |T|=k \nonumber \\
& &+\sum_{\sigma_T \in \mathcal{C}(w)} F_T \quad \quad |T|=2
\end{eqnarray}


\subsection{Equations}\label{equations_subsection}
\subsubsection{Vertices}
The Laguerre volume and surface contributions from a vertex, $v_i$,
are given by
\begin{eqnarray}
c_i S^{(i)}_i(w)& = & \Omega_i r_i^2(w) \\
c_i LIV_i^{(i)}(w) & = & \frac{\Omega_i}{3} r_i^3(w).
\end{eqnarray}
\subsubsection{Edges}
The formulas for the volume and surface contributions from the intersection of two balls $p_i$ and
$p_j$ are 

\begin{eqnarray}
c_T LIV_T^{(i)}(w) & = &\frac{\Phi_{ij}}{2} h_i^2\bigg{(}r_i(w)-\frac{h_i}{3}\bigg{)} \\
c_T LIV_T^{(j)}(w)&=&\frac{\Phi_{ij}}{2} h_j^2 \bigg{(} r_j(w) -\frac{h_j}{3}\bigg{)} \\
c_T S_T^{(i)}(w) & = & \Phi_{ij} r_i(w) h_i \\
c_T S_T^{(j)}(w) & = & \Phi_{ij} r_j(w) h_j \\
c_T LIS_T^{(i)}(w) & = & \frac{\Phi_{ij}}{2} h_i (2r_i(w)-h_i) \\
c_T LIS_T^{(j)}(w) & = & c_T LIS_T^{(i)}(w).
\end{eqnarray}

\subsubsection{Triangles}
\begin{figure}
\begin{center}
\includegraphics[scale=.2]{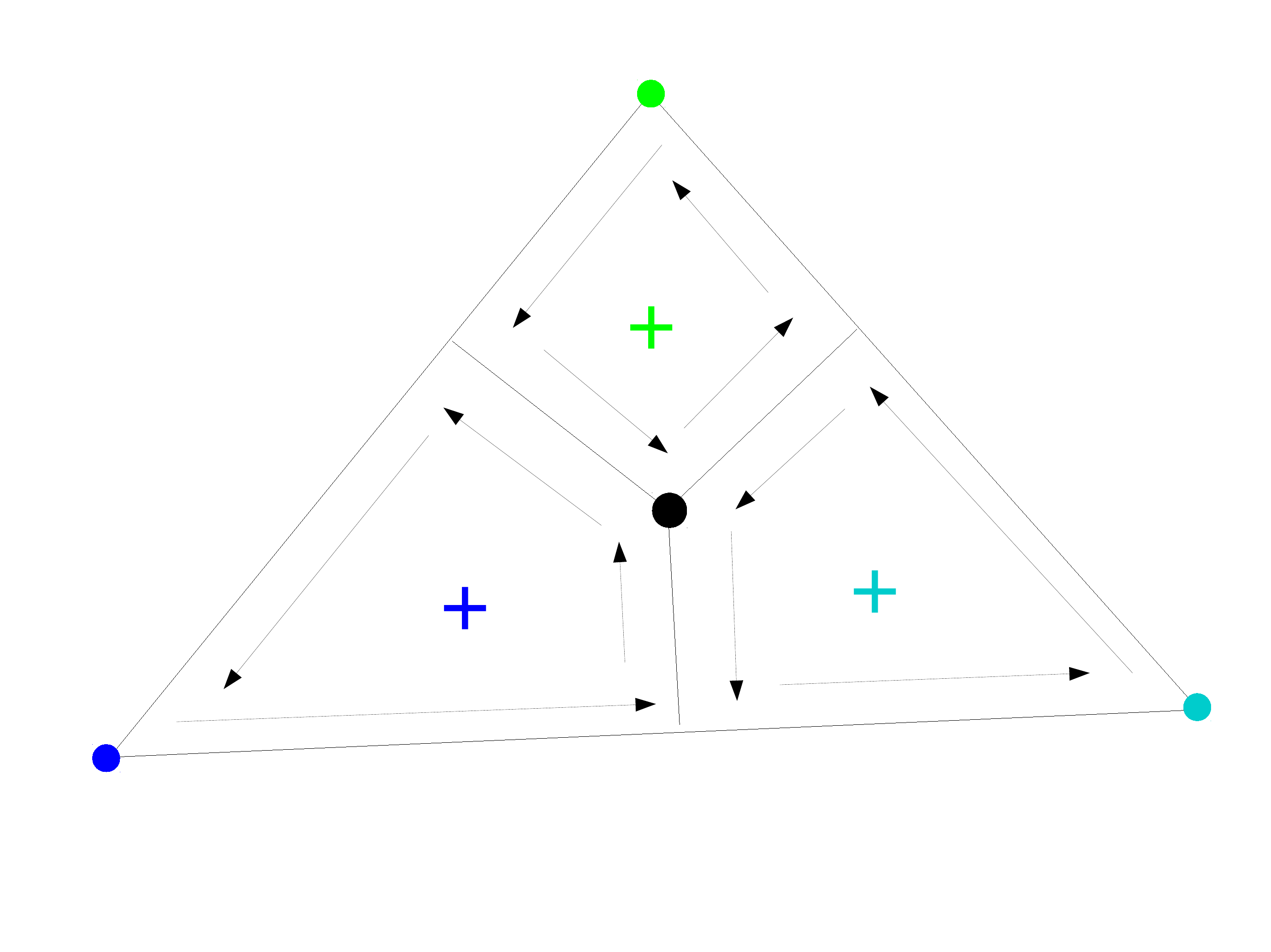}
\end{center}
\caption{$p_i'$, $p_j'$, $p_k'$, and $x_{ijk}'$ represented by dark blue, light blue, green, and black dots,
respectively. The point $p$ is out of the page. The black arrows represent positive contributions to the
volumes $\widetilde{V_i}$, $\widetilde{V_j}$,
and $\widetilde{V_k}$.}
\label{config1}
\end{figure}

\begin{figure}
\begin{center}
\includegraphics[scale=.25]{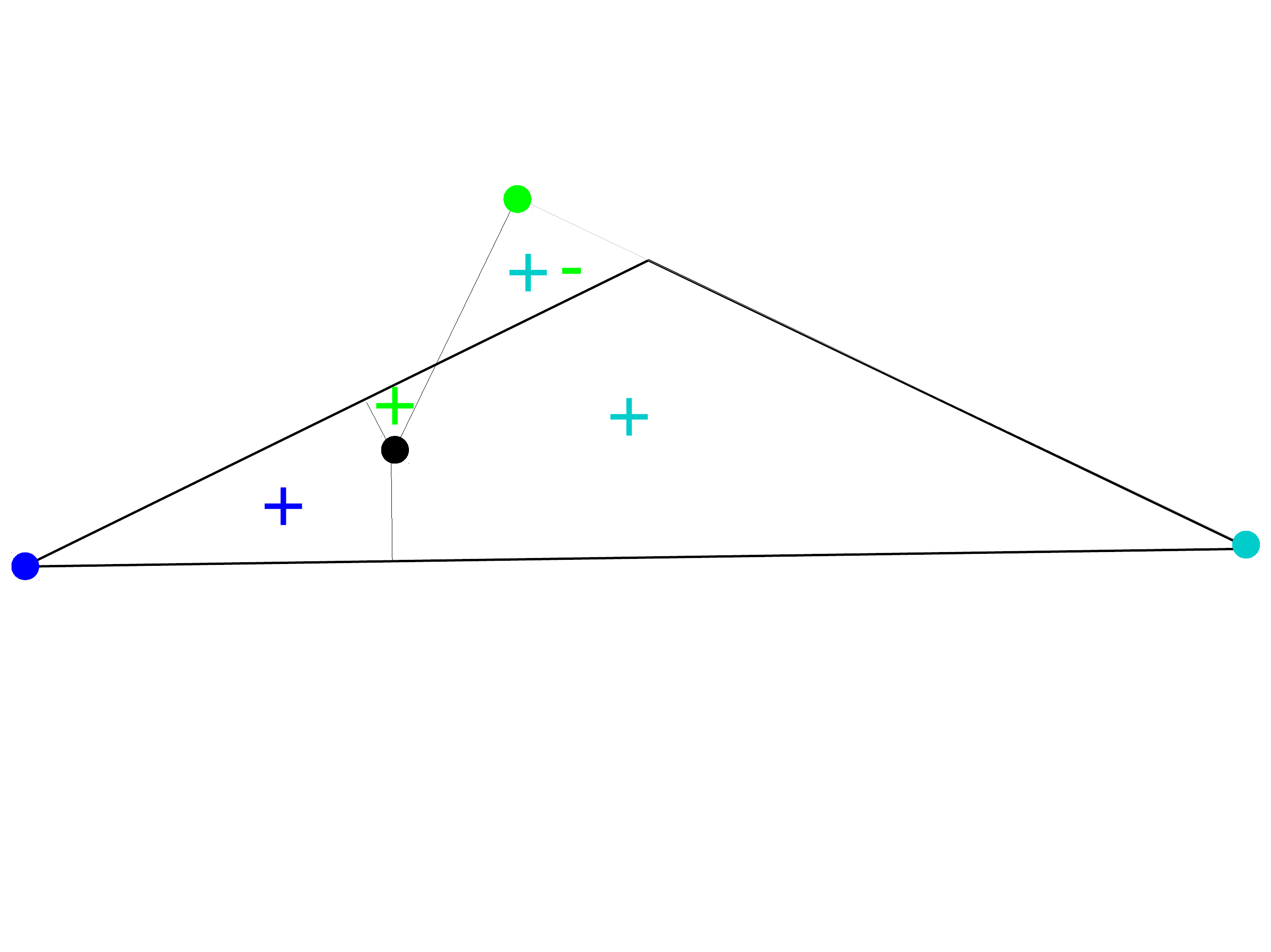}
\includegraphics[scale=.25]{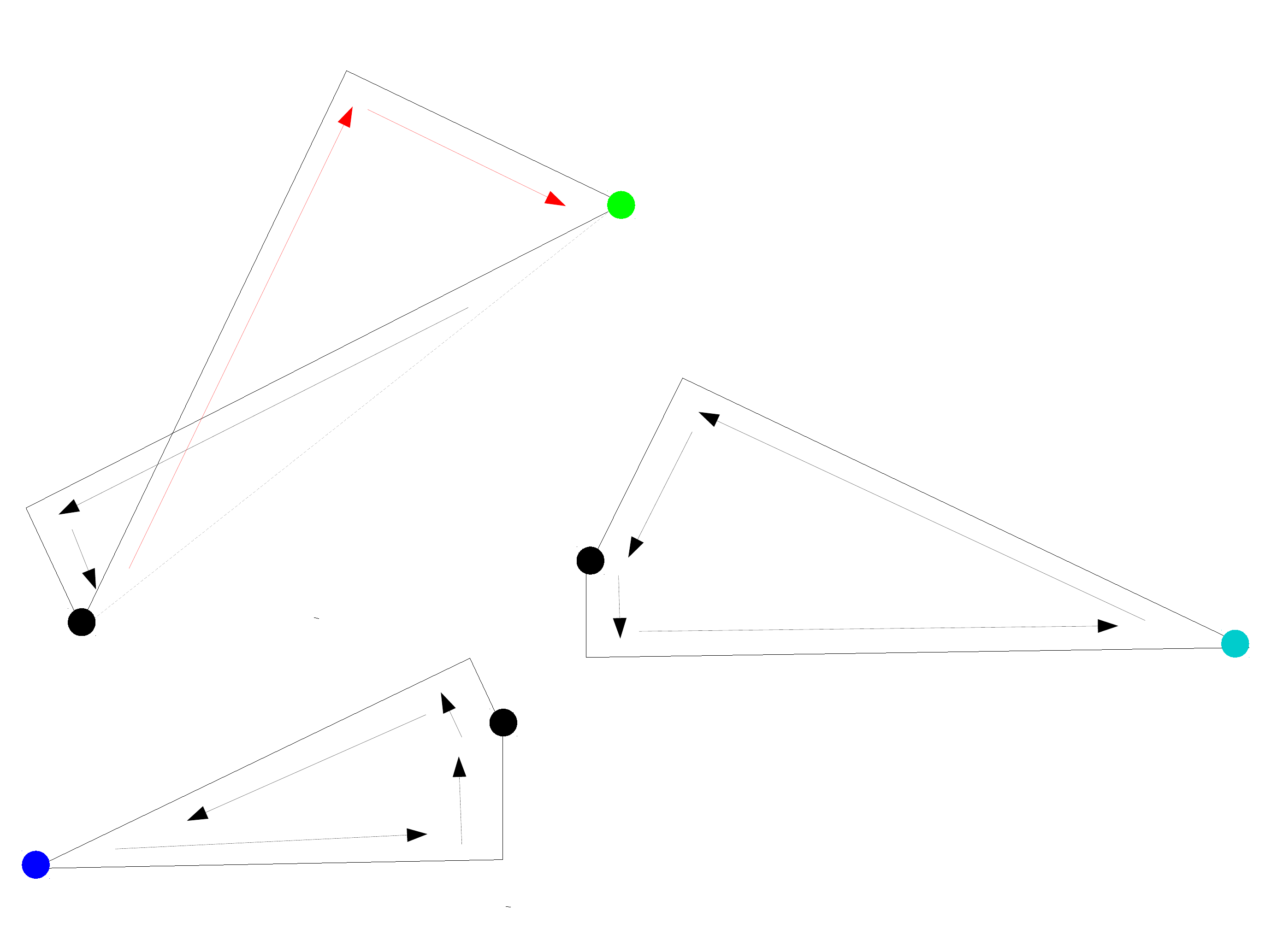}
\end{center}
\caption{Above: Sample configuration. Below: Black arrows represent positive contributions to $\widetilde{V_i}$, $\widetilde{V_j}$,
and $\widetilde{V_k}$, while red arrows represent negative contributions.}
\label{config2}
\end{figure}

\begin{figure}
\begin{center}
\includegraphics[scale=.25]{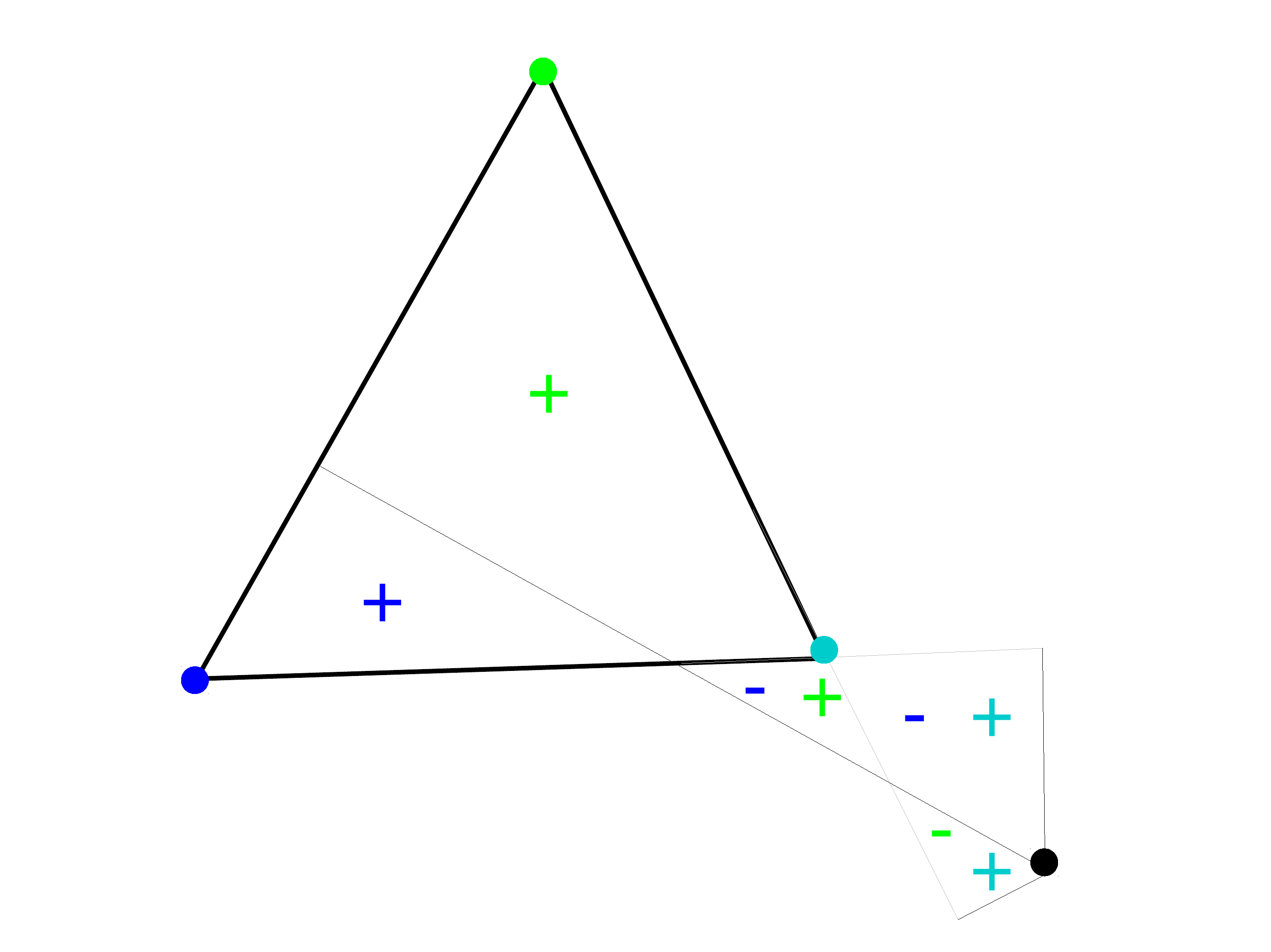}
\includegraphics[scale=.25]{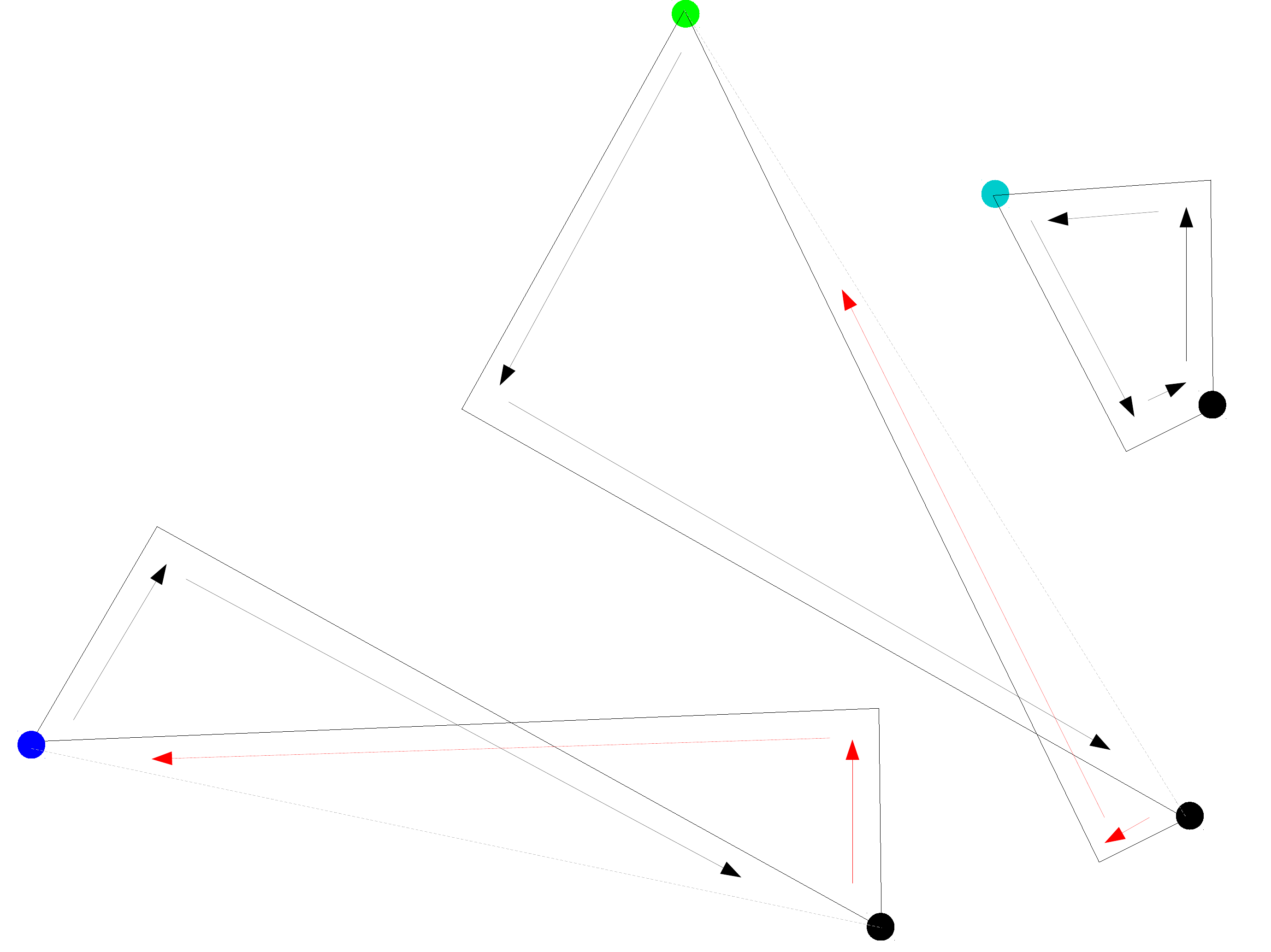}
\end{center}
\caption{Above: Sample configuration. Below: Black arrows represent positive contributions to $\widetilde{V_i}$, $\widetilde{V_j}$,
and $\widetilde{V_k}$, while red arrows represent negative contributions.}
\label{config3}
\end{figure}

Let $x_T=x_{ijk}=(x_{ijk}',x''_{ijk})$ be the characteristic point of the triangle $t_{ijk}$.
Let $p$ be one of the two points of intersection on the spheres
$p_i(w)$, $p_j(w)$, and $p_k(w)$. We can write $p=x_{ijk}'+\mathbf{n} x''_{ijk}$ where $\mathbf{n}$
is the normal to the plane containing the triangle,
\begin{equation}
\mathbf{n}= \frac{(p_j'-p_i') \times (p_k'-p_i')}{|(p_j'-p_i') \times (p_k'-p_i')|}.
\end{equation}
Let $\sigma_{Tc}$ the tetrahedron defined
by the centers of $T$ and $x_T$.
Then
\begin{eqnarray}
\frac{1}{2}S_T^{(i)} & = & \phi_{ij}S^{(i)}_{ij}+\phi_{ik}S^{(i)}_{ik}-\omega_{i} S^{(i)}_i\\
\frac{1}{2}S_T^{(j)} & = &\phi_{jk}S^{(j)}_{jk}+\phi_{ji}S^{(j)}_{ji}-\omega_{j} S^{(j)}_j\\
\frac{1}{2} S_T^{(k)} & = &\phi_{ki}S^{(k)}_{ki}+\phi_{kj}S^{(k)}_{kj}-\omega_{k} S^{(k)}_k
\end{eqnarray}
where $\phi_{ij}$ is the fractional inner dihedral angle of $\sigma_{Tc}$ along edge $\sigma_{ij}$,
$\omega_i$ is the fractional (inner) solid angle of $\sigma_{Tc}$ subtended from $p'_i$, etc.
Let $x_{ij}'$, $x_{ik}'$, $x_{jk}'$ be the centers of the characteristic
points of the triangle's edges.
Then $\frac{1}{2} LS_{ijk}^{(i)}$ is the sum of the areas of the triangles given
by $x'_{ijk}$, $x'_{ij}$, $p$, and
$x'_{ijk}$, $x'_{ik}$, $p$.
We have
\begin{equation}
\frac{1}{2}LIV^{(i)}_T = \widetilde{V_i}- \omega_i V_i^{(i)}+ \phi_{ij} LIV_{ij}^{(i)} +
\phi_{ik} LIV_{ik}^{(i)}
\end{equation}
where $\widetilde{V_i}$ is the signed volume of the solid with vertices at $p_i'$,
$x_{ij}$, $x_{ijk}$, $x_{ik}$,
and $p$. More specifically, let
\begin{eqnarray}
\mathbf{a} &=& x_{ij}'-p_i' \\
\mathbf{b} &=& x_{ijk}'-x_{ij}' \\
\mathbf{c} &=& x_{ik}'-x_{ijk}' \\
\mathbf{d} &=& p_i'-x_{ik}' \\
\end{eqnarray}
then
\begin{equation}
2 \widetilde{V_i} = \frac{h}{3} \bigg( (\mathbf{a} \times \mathbf{b})+
(\mathbf{c} \times \mathbf{d}) \bigg) \cdot \mathbf{n}
\end{equation}
Equivalent equations hold for $LIV_T^{(j)}$, $LIV_T^{(k)}$, $\widetilde{V_j}$, and $\widetilde{V_k}$.
See Figs. \ref{config1}, \ref{config2}, \ref{config3} for possible configurations.
\subsubsection{Other contributions}
\begin{figure}
\begin{center}
\includegraphics[scale=.2]{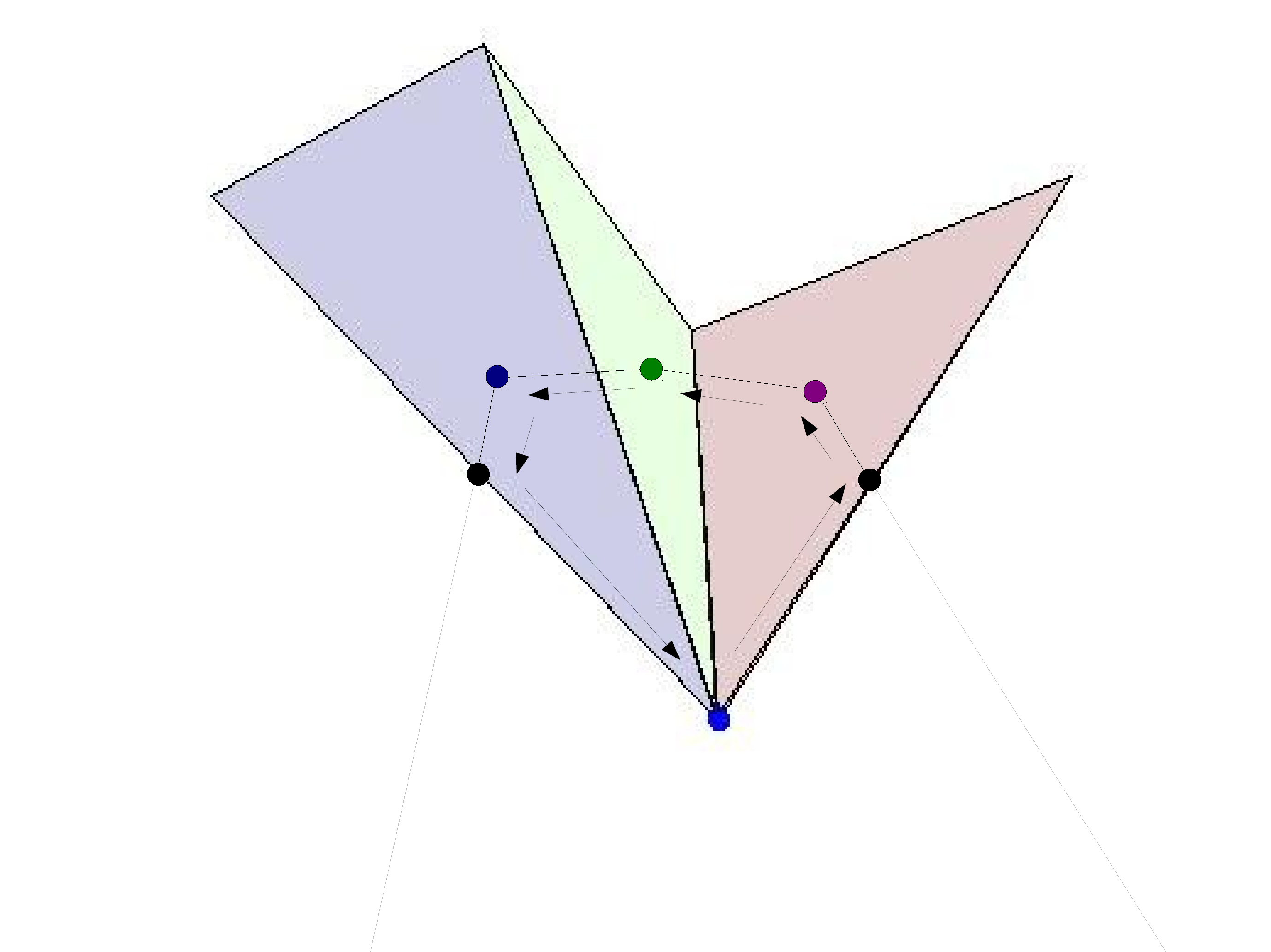}
\includegraphics[scale=.2]{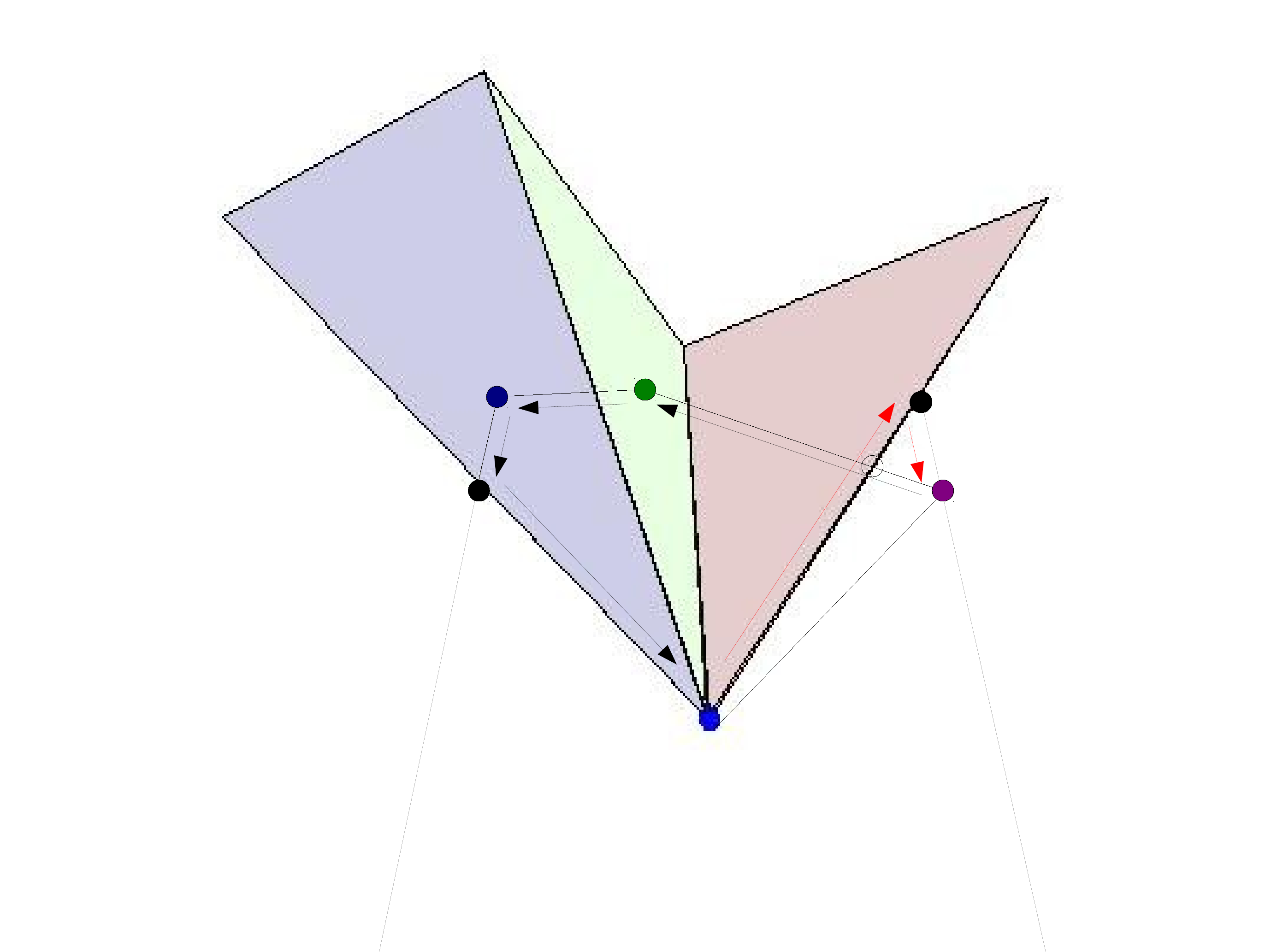}
\end{center}
\caption[Examples of additional Laguerre-Intersection volume and surface contributions
from edges in $\partial \mathcal{C}(w)$.]{Two examples of additional contributions from edges which are regular and exterior with
respect to $\mathcal{C}(w)$. The royal blue points are the characteristic points of edges
which are exterior to $\mathcal{C}(w)$ and which points directly out of the paper. The
triangles represent tetrahedra in the tetrahedra ring which are also in $\mathcal{C}(w)$.
Dark blue, green, and purple dots represent the Laguerre nodes corresponding to the tetrahedra
of the same color in the $\mathcal{C}(w)$ tetraring. The black points are characteristic
points of the boundary triangles. In the first case, both triangle characteristic points
lie in the Laguerre facet. In the second case, one of the triangle characteristic points
is not in the facet but is still needed to determine a line segment (shown in gray). In the first case
all Laguerre nodes lie in the union of tetrahedra and all corresponding surface contributions
are positive. In the second case, a Laguerre node lies outside the union of tetrahedra in
$\mathcal{C}(w)$. The triangular surface that is bounded by the right black, purple, and clear
dots was originally assigned during contributions from the boundary triangle, and must be
subtracted to obtain the correct area.}
\label{edge_reg_ext}
\end{figure}

We get additional contributions, $F_T$ and $P_T^{(i)}$, from all edges
in $\mathcal{C}(w)$. There are two types of these edges:
\begin{itemize}
\item edges interior to $\mathcal{C}(w)$
\item edges exterior to $\mathcal{C}(w)$.
\end{itemize}
Contributions from interior edges of $\mathcal{C}(w)$ are computed as in
section \ref{Interior_cells}.
That is
\begin{equation}
P_{ij}^{(i)} = LV_{ij}^{(i)}
\end{equation}
and
\begin{equation}
F_{ij}=LS_{ij}.
\end{equation}
The terms $P_T^{(i)}$ and $F_T$ are zero for all edges that are not part of at least
one tetrahedron in $\mathcal{C}(w)$. To calculate the contribution from all other
exterior $\partial \mathcal{C}(w)$ edges, we must know the characteristic points of the
triangles on the boundary of the $\mathcal{C}(w)$ tetraring. The characteristic point of the boundary
triangle may or may not be on the Laguerre facet, but it always indicates the direction of one of the
facet's boundary segments (Fig. \ref{edge_reg_ext}). For each exterior edge, the counterclockwise list of Laguerre nodes
corresponding to tetrahedra in the alpha complex is already known. The surface area is computed by
using signed surface areas as shown in pseudocode in the appendix. The volume contributions to
$LI_i$ and $LI_j$
are found by multiplying the resulting surface by $\frac{h_{ij}^i}{3}$ and $\frac{h_{ij}^j}{3}$, respectively. 
In this case, a sign is Incorporated into the surface area which means the $sign(1-t)$
and $sign(t)$ terms in equation \ref{signed_volume_equation} are not needed.

Note that the $\mathcal{C}(w)$ tetraring of an edge may not be connected. The pseudocode does
not take this into account for simplicity reasons, while the actual implementation does.
\section{Experiments: Accuracy of Algorithm}
\subsection{Monte-Carlo based estimates}
We tested the accuracy of our Laguerre-intersection algorithm on a tryptophan dipeptide trajectory (1000 structures)
with Monte-Carlo based estimates of Laguerre-Intersection volumes and surfaces.  
Tryptophan has a near-planar ring which was useful for testing the robustness of our algorithm.

Monte-Carlo Laguerre-Intersection volume estimates were obtained by generating $N=10^n$ with
$n = 1:7$ uniformly random points in the volume of each atom.
The fraction $f_{ij}$ of points in the volume of atom $j$ in structure $i$ which are closest to atom $j$ by the power distance were tabulated.
The Laguerre-Intersection volume estimate is
\begin{equation}
 \widehat{LIV_{ij}} = f_{ij} \frac{4 \pi r_j^3}{3}
\end{equation}
where $r_j$ is the radius of atom $j$.

Laguerre-Intersection atomic surface areas consist of spherical (solvent-accessible) portions and planar portions that lie on discs of intersection between two spheres.  The solvent-accessible portions
were estimated in a similar manner as Laguerre-Intersection volumes.  $N=10^n$ with $n=1:7$ uniformly random points were generated on each sphere and the estimated surface areas are
\begin{equation}
 \widehat{SAS}_{ij} = f_{ij} 4 \pi r_j^2
\end{equation}
where $f_{ij}$ is the fraction of points on the surface of atom $j$ in structure $i$ that lie closest to atom $j$ by the power distance.  

The planar Laguerre-Intersection surface area for atom $j$ is structure $i$ is
\begin{equation}
 pLIS_{ij} = \sum_{k=1}^q LIS_{ijk}
\end{equation}
where $q$ is the number of alpha-complex neighbors and 
$LIS_{ijk}$ is the portion of $LIS_{ij}$ that lies on the disc of intersection between atom $j$ and its neighbor $k$ in structure $i$.
Let $A_ijk$ be the surface area of this disc and define 
\begin{equation}
 A_{ij} = \sum_{k=1}^q A_{ijk}.
\end{equation}
Then $M_{ij} = N \frac{A_{ijk}}{A_{ij}}$ random points are generated on the disc of intersection between atom $j$ and its neighbor $k$ in structure $i$ and the estimated 
planar Laguerre-Intersection surface area is 
\begin{equation}
 \widehat{pLIS_{ij}} = \sum_{k=1}^q f_{ijk}A_{ijk}
\end{equation}
where $f_{ijk}$ is the fraction of points on the disc of intersection that are closest to atom $j$ by the power distance.

\begin{figure}[h!]
 \begin{center}
  \includegraphics[scale=.25]{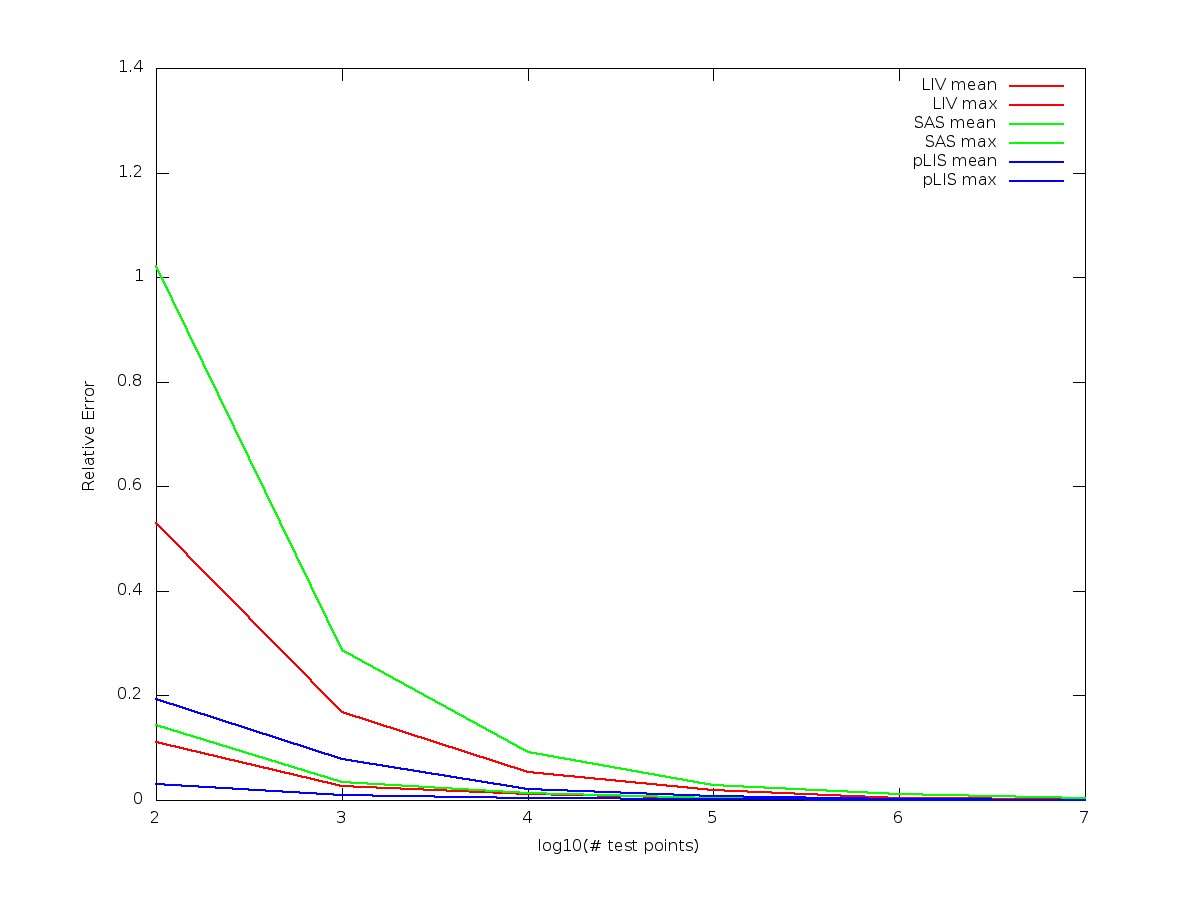}
  \caption{Mean and maximum relative difference over all atoms between estimated and analytic
  Laguerre-Intersection
  atomic volumes and surfaces areas for 1000 structure tryptophan dipeptide trajectory.} \label{rel_errors_pt_cloud}
 \end{center}
\end{figure}

Maximum and mean errors decrease as the number of test points increase
(Fig. \ref{rel_errors_pt_cloud}).
Final mean errors for $LIV$, $SAS$, and $pLIS$ estimates are $O(10^{-4})$.  Maximum relative errors
for $LIV$, $SAS$, and $pLIS$ estimates are $O(10^{-3})$.

An analysis of the accuracy of the estimates follows. Let $LIV$ be the Laguerre-Intersection volume for an atom with radius $r$.  Then the probability that 
a test point $x$ falls in $LIV$ is
\begin{equation}
 p = \frac{LIV}{4/3\pi r^3}.
\end{equation}
Given $n$ trials, the number of points, $k$, that fall within $LIV$ follows a Binomial distrubution
with paramters $p$ and $n$,
\begin{equation}
 k \sim B(p,n).
\end{equation}
For large $np$ and $n(1-p)$, $B(p,n)$ is approximated well by the normal distribution
$N(np,\sqrt{np(1-p)})$.  

Given a significance level, $\alpha = .01$, upper and lower limits, $\overline{k}$ and $\underline{k}$,
on $k$ may be calculated with $99\%$ confidence.  Since the normal distribution is 
symmetric about the mean, we are $99\%$ confident that the relative error of the estimate 
is bounded by 
\begin{equation}\label{pt_cloud_err_est}
\left|\frac{\widehat{LIV}-LIV}{LIV}\right| \leq \left|\frac{\frac{\overline{k}}{n}-p}{p}\right|.
\end{equation}
Note that the right hand side of (\ref{pt_cloud_err_est}) is a decreasing function of $p$.

We follow the same analysis for $SAS$ and $pLIS$.  
For our trajectory, 
$p\geq .174$, $p \geq .056$, and $p\geq .180$ for $LIV$, $SAS$, and $pLIS$ respectively.  This gives
99\% probability that 
\begin{eqnarray}
 \left|\frac{\widehat{LIV}-LIV}{LIV}\right|& \leq &1.77 \times 10^{-3} \\
 \left|\frac{\widehat{SAS}-SAS}{SAS}\right|& \leq &3.34 \times 10^{-3} \\
 \left|\frac{\widehat{pLIS}-pLIS}{pLIS}\right|& \leq &1.73 \times 10^{-3}
\end{eqnarray}
which is consistent with the values in Fig. \ref{rel_errors_pt_cloud}. 

\subsection{Comparison with Gauss-Bonnet-type algorithm}
We compared atomic Laguerre-Intersection volumes and surfaces areas
calculated using our method with those found using
Gauss-Bonnet-type algorithm
adopted in the SBL library on 411 proteins in the Protein Data Bank.  With the 
latter algorithm, we used the 
'exact' option which employs interval arithmetic with sufficiently small 
intervals that the solution is considered 'exact'.  Average 
surface area error per atom was $2.266\times10^{-13}$ while average volume error
per atom was $4.333\times 10^{-13}$.  The maximum surface area error over 
all atoms was $1.188\times 10^{-7}$ and maximum volume error over all atoms was 
$6.468\times 10^{-8}$.  Average runtime for our algorithm was .67 seconds compared 
to 35.06 seconds for the interval arithmetic algorithm. While on average, our calculations agreed with the exact 
calculations to 13 digits, our method was about 50 times faster.

\section{Experiments: Optimizing Solvent Parameter}
We test the capability of the Laguerre-Intersection method for predicting
explicit water Laguerre quantities.  We train on molecular dynamics trajectory data to
determine ``optimal solvent parameters''; those parameters for which the Laguerre-Intersection 
volumes and surface areas are, on average, as close as possible to the Laguerre volumes and surfaces of
a molecule in its typical solvent environment.

\subsection{Data}\label{optimizing_section}

\begin{figure}
\begin{center}
\includegraphics[scale=.35]{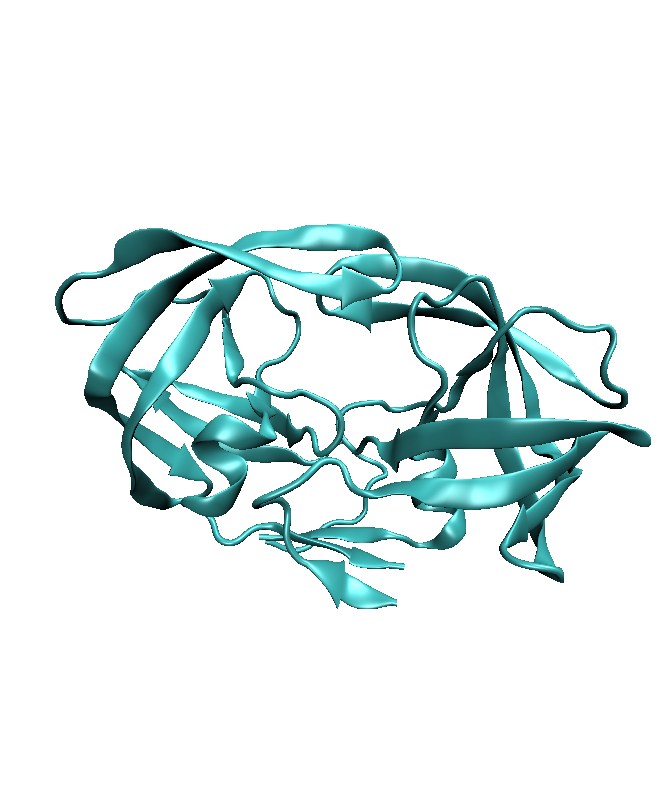}
\end{center}
\caption{Sample structure from the HIV protease molecular dynamics trajectory.}\label{HIV_snapshot}
\end{figure}

We work with solvated molecular dynamics trajectory data (2500 structures) of the HIV protease dimer
(See Fig. \ref{HIV_snapshot}) to determine optimal solvent parameters.  The trajectory was initiated 
from the 1HVR crystal structure \cite{Lam:1994}, and generated 
using the Amber software \cite{Case:2005} with the ff99SB protein force field \cite{Hornak:2006}
and the TIP3p water model \cite{Jorgensen:1983}. Equilibration followed published protocol 
\cite{Shang:2012} with a 2fs time step, and the simulation coupled to a thermostat at 300K.  A total of 
100 ns of data were generated. 

Optimal solvent parameters, which 
converge quickly, determine the capability of the implicit Laguerre-Intersection model 
to predict explicit water Laguerre quantities.  We also compare our method to other implicit models.

First residual and atomic Laguerre volumes, atomic, interresidual, 
and residue-solvent surface areas (LV\_res, LV\_atom, LS\_atom, LS\_interres, LSAS\_res, respectively) were calculated 
for the solvated trajectory.

For various solvent parameters, $w_k$
and $r_k$,
\begin{eqnarray}
w_k & = & k \cdot dw \nonumber \\
r_k & = & k \cdot dr \nonumber
\end{eqnarray}
equivalent Laguerre-Intersection quantities were computed for each structure (LIV\_res, LIV\_atom,
LIS\_atom, LIS\_interres, LISAS\_res, respectively).  We set $dw=.1$ \AA\ and chose $dr$ such that 
\begin{equation}
1.7^2+k_{max}\cdot dw = (1.7 + k_{max}\cdot dr)^2.
\end{equation} 
This means that for an atom with 
atomic radius $1.7$ \AA\ (the typical radius of carbon), the initial ($k=0$) and final ($k=k\_max$)
total weights are the same using both methods. 

Laguerre and
Laguerre-Intersection quantities were compared and ``optimal'' solvent parameters were found.
The optimal solvent weight (radius) is the value $w_k$ ($r_k$) that
minimizes the error (Equations \ref{1norm}, \ref{2norm}).

We write the Laguerre volume of residue $j$ in structure $i$ as $LV\_res(i,j)$, and the Laguerre-Intersection
volume of residue $j$ in structure $i$ with solvent parameter $w_k$ ($r_k$) as $LIV\_res(i,j,k)$.  
We write the interresidual Laguerre surface area between residues $j$ and $jj$ in structure $i$ as
$LS\_interres(i,j,jj)$, and the interresidual Laguerre-Intersection
surface area between residues $j$ and $jj$ in structure $i$ with solvent value $w_k$ ($r_k$) as
$LIS\_interres(i,j,jj,k)$.  Equivalent formulas hold for the other Laguerre and 
Laguerre-Intersection quantities.

Using wildcard notation, an arbitrary Laguerre quantity and Laguerre-Intersection quantity
are written as $L^*$ and $LI^*$.  Then with solvent parameter $w_k$ ($r_k$) 
 1-Norm and 2-Norm errors are
\begin{equation}\label{1norm}
 E_1(LI^*(k)) = \frac{1}{N} \sum_{i=1}^{N} \epsilon_{1,i}(LI^*(k))
\end{equation}
and
\begin{equation}\label{2norm}
 E_2(LI^*(k)) = \sqrt{\frac{1}{N} \sum_{i=1}^{N} \epsilon_{2,i}(LI^*(k))}
\end{equation}
where $N$ is the number of structures in the MD trajectory.

For $LIV\_res$, $LIV\_atom$, $LIS\_atom$, and $LISAS\_res$
\begin{equation}
 \epsilon_{1,i}(LI^*(k)) = \frac{1}{\overline{n}}\sum_{j=1}^{n} |LI^*(i,j,k)-L^*(i,j)|
\end{equation}
and
\begin{equation}
 \epsilon_{2,i}(LI^*(k)) = \frac{1}{\overline{n}}\sum_{j=1}^{n} (LI^*(i,j,k)-L^*(i,j))^2
\end{equation}
where $n$ is the number of residues or atoms and

\begin{equation}
 \overline{n}= \sum_{j=1}^{n} \mathbb{I}(j)
\end{equation}
with

 \begin{displaymath}
   \mathbb{I}(j) = \left\{
     \begin{array}{lr}
       1 & \text{if } LI^*(i,j) \neq 0 \text{ or } L^*(i,j) \neq 0\\
       0 & \text{otherwise}
     \end{array}
   \right.
\end{displaymath} 
For the first three quantities $\overline{n}=n$.  The average Laguerre quantity is 
defined as 
\begin{equation}\label{average_Laguerre}
 av(L^*) = \frac{1}{N} \sum_{i=1}^{N} \Bigg(\frac{1}{\overline{n}}\sum_{j=1}^{n} L^*(i,j)\Bigg).
\end{equation}

With $LIS\_interres$ we have
\begin{equation}
 \epsilon_{1,i}(LI^*(k)) = \frac{1}{\overline{n}}\sum_{j=1}^{n} \sum_{jj=j+1}^{n}|LI^*(i,j,jj,k)-L^*(i,j,jj)|
\end{equation}
and
\begin{equation}
 \epsilon_{2,i}(LI^*(k)) = \frac{1}{\overline{n}}\sum_{j=1}^{n} \sum_{jj=j+1}^{n} (LI^*(i,j,jj,k)-L^*(i,j,jj))^2
\end{equation}
where 
\begin{equation}
 \overline{n}= \sum_{j=1}^{n} \sum_{jj=j+1}^{n}\mathbb{I}(j,jj)
\end{equation}
with

 \begin{displaymath}
   \mathbb{I}(j,jj) = \left\{
     \begin{array}{lr}
       1 & \text{if } LI^*(i,j,jj) \neq 0 \text{ or } L^*(i,j,jj) \neq 0\\
       0 & \text{otherwise.}
     \end{array}
   \right.
\end{displaymath} 
 The average interresidual surface area is defined as
\begin{equation}\label{average_Laguerre_LS_interres}
  av(LS\_interres) = \frac{1}{N} \sum_{i=1}^N 
  \Bigg( \frac{1}{\overline{n}} \sum_{j=1}^n \sum_{jj=j+1}^n L^*(i,j,jj) \Bigg).
\end{equation}

In this paper we present the absolute errors divided by the average quantities.  We do not 
compute the relative errors at each step because the exact quantities may be zero.

First we present optimal solvent parameters obtained from HIV protease trajectory data.  We show 
these parameters converge quickly despite the configurational variability of the molecule.
We compare the capability of the Laguerre-Intersection, McConkey \cite{constrained_Voronoi}, and Cazals' \cite{Cazals:2010} 
method to predict explicit solvent Laguerre quantities.  After a brief overview of these methods,
We will compare all three 
methods to explicit water 
models, followed by a comparison of the radius vs. weight method of expanding
atoms.

\subsection{Optimal solvent parameters}

\begin{figure}[h!]
 \begin{center}
  \includegraphics[scale=.07]{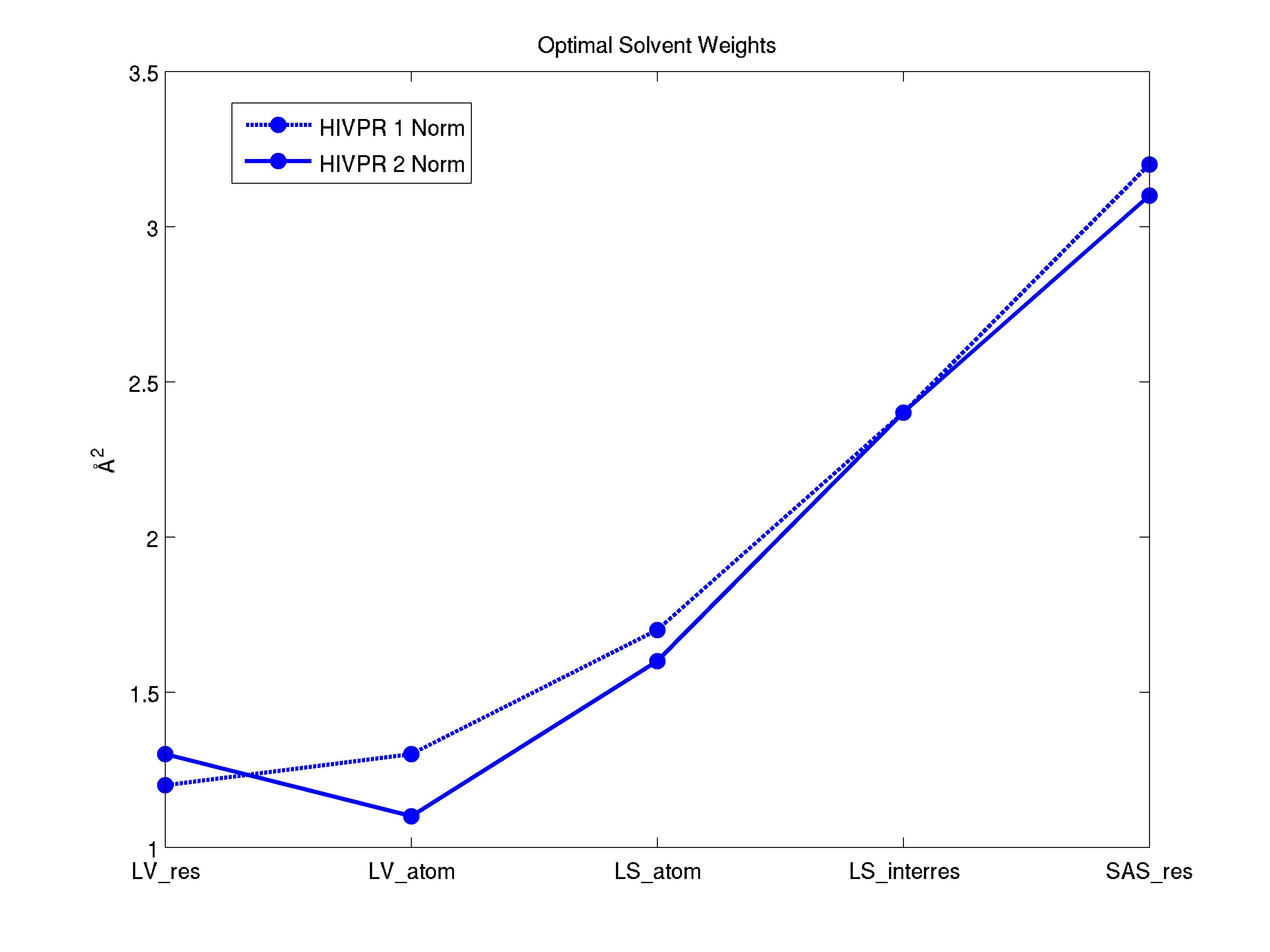}
  \caption{Optimal solvent weights from the HIV protease trajectory.} \label{optimal_parameters}
 \end{center}
\end{figure}

We plot the optimal solvent weights found from the HIV protease trajectory which has 2500 structures 
(See Fig. \ref{optimal_parameters}).
The optimal solvent parameters for Laguerre volumes and Laguerre surfaces differ which is expected.  One additional 
benefit of the weight method, is that the weighted Delaunay tetrahedrization is constant
for all weights.  This means the tetrahedrization (the time limiting step)
only needs to be computed once per structure.  When using the radius method, 
the tetrahedrization must be recalculated for each optimal solvent parameter.

Note that 
these 
are not ``universal'' parameters and are not necessarily optimal for an 
arbitrary molecule.  
Further work is needed to determine which values should be used in 
a given molecular dynamics simulation.

\begin{figure}
 \begin{center}
  \includegraphics[scale=.4]{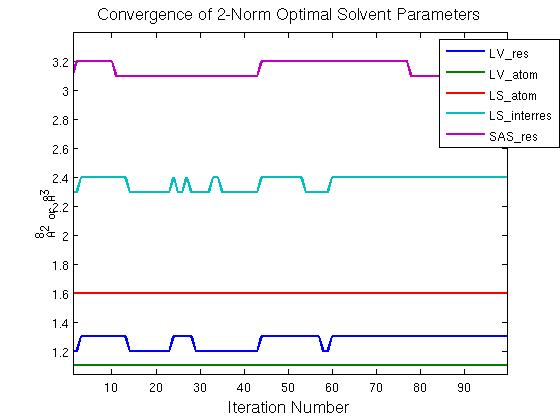}
  \caption{Convergence of optimal solvent parameters.}\label{convergence_fig}
 \end{center}
\end{figure}

Optimal solvent parameters, as given by Equations \ref{1norm}, \ref{2norm},
\ref{average_Laguerre}, and \ref{average_Laguerre_LS_interres},
  are recorded at each step in the trajectory. The first 100 out of 2500 
  iterations are shown in Fig. \ref{convergence_fig}.  We see that the 
  optimal solvent parameters are immediately within one $dw$ of the 
  global optimal solvent parameters and soon become stationary with each step.  
  Since the HIV protease dimer has flexible loops and open-close transitions we 
  conjecture that most other proteins will exhibit similar or quicker convergence.  

\subsection{Overview of current methods}	
\subsubsection{McConkey-Cazal method}\label{McConkey_section}
McConkey-Cazals (MC) algorithm \cite{Cazals:2011} \cite{constrained_Voronoi} 
constructs Laguerre-like cells by considering the surfaces of extended radical contact
planes between neighboring atoms within a cutoff distance and the surface of an
expanded sphere. Similar to our Laguerre-Intersection algorithm, 
this method considers the intersection of cells with the space filling diagram with atom radii 
expanded by a solvent radius of $1.4$ \AA.  While the McConkey-Cazals method and the Laguerre-Intersection method are algorithmically different,
 (McConkey-Cazals method is based on the Gauss-Bonnet theorem while ours is an in inclusion-exlusion method), another key difference 
is that the cells in McConkey-Cazals method are bounded by extended radical planes rather
than radical planes. This
poses problems since the cells of atoms in the bulk, which should not be affected by the cutoff distance,
are affected. Furthermore, cells of small atoms completely disappear 
for cutoff distances as small as 1.4 \AA.


\subsubsection{'Intervor' for interface surfaces}
Another Cazals' method, 'Intervor', is used to determine interface surfaces between molecules
or a molecule and solvent \cite{Cazals:2006} \cite{Cazals:2010} and locates atoms that are in direct 
contact with each other and those whose contact is mediated by water.  This method is 
more qualitative in nature than the Laguerre-Intersection method, i.e. it does 
not claim to give contact areas that are similar to those of solvated systems but rather locate and describe 
topology of interactions.  

A solvent radius of $r=1.4$ \AA\ to all atoms and the alpha complex, $\mathcal{C}(0)$, 
is computed.  Laguerre facets dual to edges not in $\mathcal{C}(0)$ are thrown out.
This is 
called the $\alpha$ criterion.  However, overly large facets still remain.  Such facets are discarded 
according to 'condition $\beta$':

\begin{equation}
 \frac{\overline{\mu}}{w_e} > M^2
\end{equation}
where $w_e$ is the weight of the smaller of the two balls in the edge and $\overline{\mu}$ is
the size of the largest weighted Delaunay tetrahedron that contains the edge.  The value $M$ is set to 
$5$.

\subsection{Comparison with explicit model}\label{comparison_explicit}
We compare the Laguerre-Intersection, McConkey-Cazals \cite{constrained_Voronoi} \cite{Cazals:2011} (MC),
and Intervor \cite{Cazals:2010} to explicit models 
for the HIV protease trajectory data. 
\begin{table}
 \begin{center}
 \begin{tabular}{|l|l|c|c|c|}
 \hline
Quantity & Method &1-Norm / Average &2-Norm / Average \\
 \hline
LS\_interres  & Laguerre-Intersection &.1416& .2027 \\
 \hline
&MC r=1.4 &.3224 &.5101\\
 \hline
&Intervor &.6988& 1.3150 \\
 \hline
 \hline
LV\_res & LI& .05695 & .07097 \\
 \hline
&MC r=1.4 &.4184& .5487 \\
 \hline
 \hline
 LS\_atom  &LI &.06794 & .09994 \\
 \hline
&MC r=1.4 & .4841& .5714  \\
  \hline
 \hline
LV\_atom &LI &.1039 & .1636\\
 \hline
&MC r=1.4 & .7573 & 1.0937 \\
 \hline 
 \hline
SAS\_res & LI& .1176 &.1548 \\
 \hline
&MC r=14. &.2039&.2536 \\
 \hline
 \end{tabular}
 \caption{Ratio of errors to average Laguerre quantities.
 See equations \ref{1norm}, \ref{2norm}, \ref{average_Laguerre}, \ref{average_Laguerre_LS_interres}.
 }
 \end{center}
 \label{errors_table}
 \end{table}
 We see a clear improvement in accuracy using the Laguerre-Intersection method (See Table 1).
 SAS\_res is approximately thirty percent more accurate while LV\_res is about 
 eight times more accurate when compared to the MC method.  Intervor\cite{Cazals:2010}
 gives norms about four times larger than the Laguerre-Intersection method and about 
 two times larger than the MC method.
 
 At this point we do not know what effect this improvement might have on the accuracy of statistical potentials 
 developed with these models.

\subsection{Radius vs. weight method}\label{radius_vs_weight}

\begin{figure}[h!]
  \includegraphics[scale=.07]{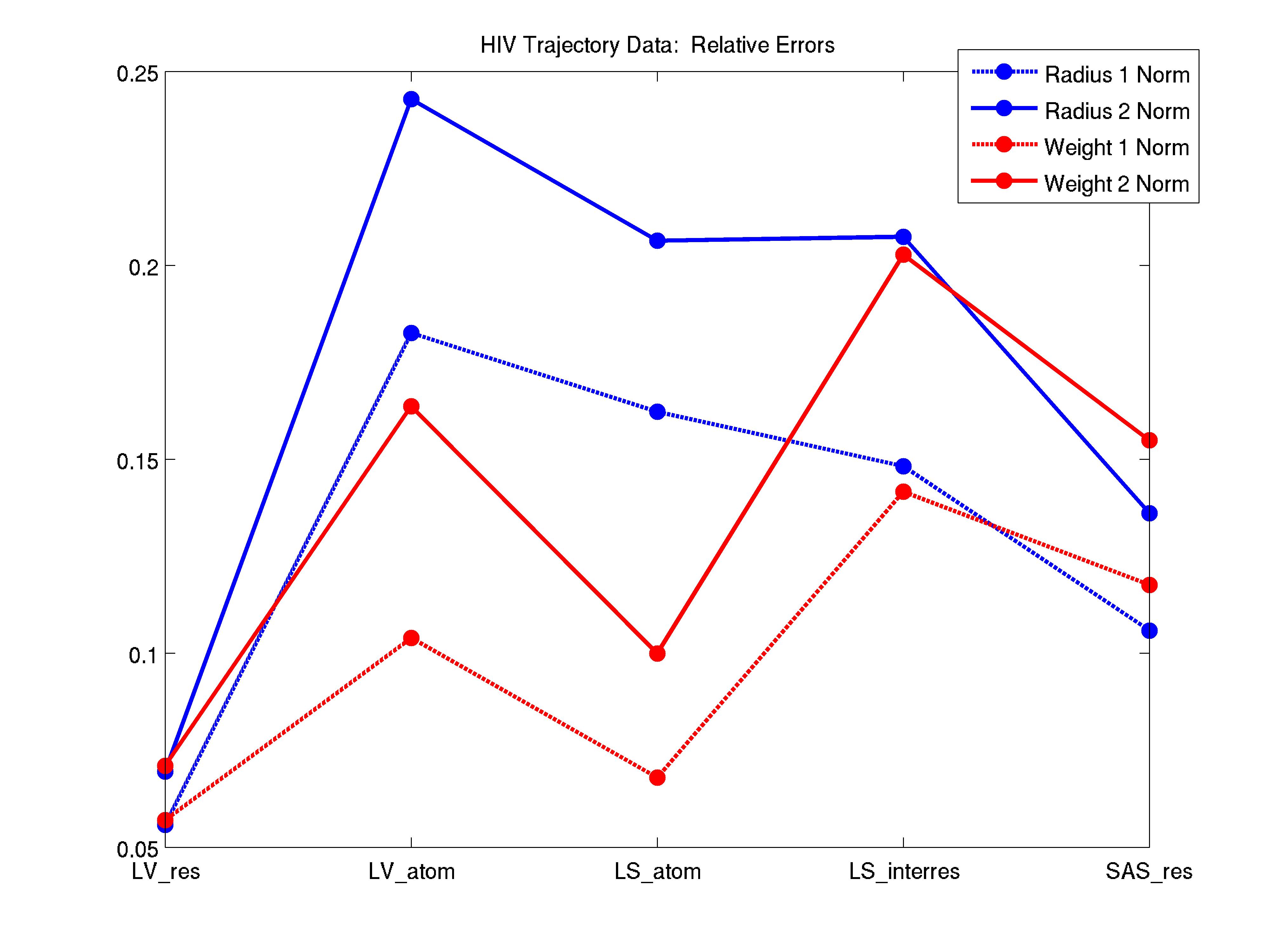}
  \caption{Relative errors corresponding to optimal solvent parameters.}
  \label{HIV_errors}
\end{figure}

In our simulations, we found that the weight rather than radius method gave smaller
errors for atomic quantities and similar errors for residual quantities.  This is expected as 
the Laguerre cell of the molecule does not vary as atom weights are uniformly increased
(See Fig. \ref{Laguerre_vary_weight}).  
Residual errors decrease relative to atomic errors for the radius method due to cancellation in atomic errors.  Fig.
\ref{HIV_errors} compares 
relative errors for Laguerre-Intersection quantities
over the HIV protease trajectory. 

\section{Conclusion}
We developed an algorithm to compute the atomic and residual volumes and surface areas of
the intersection of the Laguerre diagram with the space filling model of a molecule with
applications to implicit solvation.  While methods exist which calculate volumes and
surface areas of the restriction of the Laguerre diagram to the space filling model,
our algorithm is unique in that it is based on inclusion-exclusion rather
than Gauss-Bonnet which decreases the complexity of the calculation. 
We also optimize an adjustable parameter, the weight, so the Laguerre-Intersection atomic
and residual volumes and surface areas are as close as possible to Laguerre volumes and
surface areas found using explicit solvent.  We test our algorithm on an explicit water
HIV protease molecular dynamics trajectory.  
We show that volumes and surface areas computed using our implicit method are 30\% 
to 8 times closer to explicit quantities than those found using current models.
We also show that varying the weight rather than the radius in the space-filling
diagram gives much better agreement with explicit quantities.

\section*{Acknowledgments}
Michelle Hummel and Evangelos Coutsias were supported in part by NIH grant R01GM090205 and
Stony Brook University research funds.  Carlos Simmerling was supported in part by NIH grant R01GM107104.
Evangelos Coutsias and Carlos Simmerling acknowledge support 
from the Laufer Center for Physical and Quantitative Biology at Stony Brook University.  A 
portion of this work was carried out at the Department of Mathematics and Statistics, Univerisity 
of New Mexico.

\bibliographystyle{amsplain}
\bibliography{mybib2}

\providecommand{\bysame}{\leavevmode\hbox to3em{\hrulefill}\thinspace}
\providecommand{\MR}{\relax\ifhmode\unskip\space\fi MR }
\providecommand{\MRhref}[2]{%
  \href{http://www.ams.org/mathscinet-getitem?mr=#1}{#2}
}
\providecommand{\href}[2]{#2}
\begin{thebibliography}{10}

\bibitem{Ban:2006}
Yih-En~Andrew Ban, Herbert Edelsbrunner, and Johannes Rudolph, \emph{Interface
  surfaces for protein-protein complexes}, Journal of the ACM \textbf{53}
  (2006), 361--378.

\bibitem{Case:2005}
David~A. Case~et al., \emph{The {A}mber biomolecular simulation programs},
  Journal of Computational Chemistry \textbf{26} (2005), 1668--1688.

\bibitem{Cazals:2010}
Frederic Cazals, \emph{Revisiting the {V}oronoi description of protein-protein
  interfaces: Algorithms}, Lecture Notes in Computer Science \textbf{6282}
  (2010), 419--430.

\bibitem{Cazals:2011}
Frederic Cazals, Harshad Kanhere, and Sebastien Loriot, \emph{Computing the
  volume of a union of balls: a certified algorithm}, ACM Transactions on
  Mathematical Software \textbf{38} (2011).

\bibitem{Cazals:2006}
Frederic Cazals, Flavien Proust, Ranjit~P. Bahadur, and Joel Janin,
  \emph{Revisiting the {V}oronoi description of protein-protein interfaces},
  Protein Science \textbf{15} (2006), 2082--2092.

\bibitem{Connolly:1983}
M.~L. Connolly, \emph{Analytical molecular surface calculation}, Journal of
  applied crystallography \textbf{16} (1983), 548--558.

\bibitem{WAlpha}
Herbert Edelsbrunner, \emph{Weighted alpha shapes}, Technical Report (1992).

\bibitem{Union}
\bysame, \emph{The union of balls and its dual shape}, Proceeding of the ninth
  annual symposium on Computational Geometry, San Diego, CA, 1993.

\bibitem{SFandV}
Herbert Edelsbrunner and Ping Fu, \emph{Measuring space filling diagrams and
  voids}, Proc. 28th Ann. HICSS (1995).

\bibitem{bio_solv}
Herbert Edelsbrunner and Patrice Koehl, \emph{The geometry of biomolecular
  solvation}, Discrete and Computational Geometry: MSRI publications
  \textbf{52} (2005), 241--273.

\bibitem{Residue_vols}
Jeremy Esque, Christophe Oguey, and Alexandre~G. de~Brevern, \emph{A novel
  evaluation of residue and protein volumes by means of {L}aguerre
  tessellation}, J. Chem. Inf. Model \textbf{50} (2010), 947--960.

\bibitem{Laguerre_protein_contacts}
\bysame, \emph{Comparative analysis of threshold and tessellation methods for
  determining protein contacts}, Journal of Chemical Information and Modeling
  \textbf{51} (2011), 493--507.

\bibitem{Volume_atoms_protein_surf_Voronoi}
Mark Gerstein, Jerry Tsai, and Michael Levitt, \emph{The volume of atoms on the
  protein surface: Calculated from simulation, using {V}oronoi polyhedra}, J.
  Mol. Biol. \textbf{249} (1995), 955--966.

\bibitem{Vol3}
K.D. Gibson and Harold~A. Scheraga, \emph{Volume of the intersection of three
  spheres of unequal size: a simplified formula}, Journal of Physical Chemistry
  \textbf{91} (1987), 4121--4122.

\bibitem{Headd:2007}
Jeffrey~J. Headd, Y.E.~Andrew Ban, Paul Brown, Herbert Edelsbrunner, Madhuwanti
  Vaidya, and Johannes Rudolph, \emph{Protein-protein interfaces: Properties,
  preferences, and projections}, Journal of Proteome Research \textbf{6}
  (2007), 2576--2586.

\bibitem{Hornak:2005}
Viktor Hornak~et al., \emph{Hiv-1 protease flaps spontaneously open and reclose
  in molecular dynamics simulations}, PNAS \textbf{103} (2005), 915--920.

\bibitem{Hornak:2006}
\bysame, \emph{Comparison of multiple {A}mber force fiels and development of
  improved protein backbone parameters}, Proteins: Structure, Function, and
  Bioinformatics \textbf{65} (2006), 712--725.

\bibitem{Hummel:2014}
Michelle~H. Hummel, \emph{{Delaunay-Laguerre Geometry for Macromolecular
  Modeling and Implicit Solvation}}, Ph.D. thesis, University of New Mexico,
  New Mexico, December 2014, Lobovault Repository,
  http://hdl.handle.net/1928/25782.

\bibitem{Jorgensen:1983}
William~L. Jorgensen~et al., \emph{Comparison of simple potential functions for
  simulating liquid water}, The journal of chemical physics \textbf{79} (1983),
  926--935.

\bibitem{Kollman:2000}
Peter~A. Kollman~et al., \emph{Calculating structures and free energies of
  complex molecules: combining molecular mechanics and continuum models}, Acc.
  Chem. Res. \textbf{33} (2000), 889--897.

\bibitem{Lam:1994}
PY~Lam~et al., \emph{Rational design of potent, bioavailable, nonpeptide cyclic
  ureas as hiv protease inhibitors}, Science \textbf{263} (1994), 380--384.

\bibitem{ShapeComp}
Jie Liang, Herbert Edelsbrunner, Ping Fu, Pamidighantam~V. Sudhaker, and
  Shankar Subramaniam, \emph{Analytical shape computation of macromolecules: I.
  molecular area and volume through alpha shapes}, PROTEINS: Structure,
  Function, and Genetics \textbf{33} (1998), 1--17.

\bibitem{Mahdavi:2012}
Sedigheh Mahdavi, Ali Mohades, Ali Salehzadeh-Yazdi, Samad Jahandideh, and Ali
  Masoudi-Nejad, \emph{Computational analysis of {RNA}-protein interaction
  interfaces via the {V}oronoi diagram}, Journal of Theoretical Biology
  \textbf{293} (2012), 55--64.

\bibitem{Mahdavi:2013}
Sedigheh Mahdavi, Ali Salehzadeh-Yazdi, Ali Mohades, and Ali Masoudi-Nejad,
  \emph{Computational structure analysis of biomacromolecule complexes by
  interface geometry}, Computational Biology and Chemistry \textbf{47} (2013),
  16--23.

\bibitem{constrained_Voronoi}
B.J. McConkey, V.~Sobolev, and M.~Edelman, \emph{Quantification of protein
  surfaces, volumes and atom-atom contacts using a constrained {V}oronoi
  procedure}, Bioinformatics \textbf{18} (2002), 1365--1373.

\bibitem{Nguyen:2014}
Hai Nguyen~et al., \emph{Folding simulations for proteins with diverse
  topologies are accessible in days with a physics-based force field and
  implicit solvent}, Journal of the American Chemical Society \textbf{136(40)}
  (2014), 13959--13962.

\bibitem{CAD_score}
Kliment Olechnovic, Eleonora Kulberkyte, and Ceslovas Venclovas,
  \emph{Cad-score: A new contact area difference-based function for evaluation
  of protein structural models}, Proteins \textbf{81} (2013), 149--162.

\bibitem{dynamic_interface_between_proteins}
N.~Ray, X.~Cavier, J.C. Paul, and B.~Maigret, \emph{Intersurf: dynamic
  interface between proteins}, J Mol Graph Model \textbf{23} (2005), 347--354.

\bibitem{interface}
Nicolas Ray, Xavier Cavin, and Bernard Maigret, \emph{Interactive poster:
  visualizing the interaction between two proteins}, Proceedings, IEEE
  Visualization 2003 (2003).

\bibitem{Lag_decomp_app_protein_folds}
J.F. Sadoc, R.~Jullien, and N.~Rivier, \emph{The {L}aguerre polyhedral
  decomposition: application to protein folds}, The European Physical Journal B
  \textbf{33} (2003), 355--363.

\bibitem{Shang:2012}
Yi~Shang and Carlos Simmerling, \emph{Molecular dynamics applied in drug
  discovery: The case of hiv-1 protease}, Methods in Molecular Biology
  \textbf{819} (2012), 527--549.

\bibitem{surf_vs_bulk_Voronoi}
Alain Soyer, Jaques Chomilier, Jean-Paul Mornon, Remi Jullien, and
  Jean-Francois Sadoc, \emph{{V}oronoi tessellation reveals the condensed
  matter character of folded proteins}, Physical Review Letters \textbf{85}
  (2000), 3532--3535.

\bibitem{Tan:2007}
Chunhu Tan, Yu-Hong Tan, and Ray Luo, \emph{Implicit nonpolar solvent models},
  The journal of physical chemistry \textbf{111} (2007), 12263--12274.

\bibitem{Weiser:1999}
Jorg Weiser, Peter~S. Shenkin, and W.~Clark Still, \emph{Approximate atomic
  surfaces from linear combinations of pairwise overlaps (lcpo)}, Journal of
  Computational Chemistry \textbf{20} (1999), 217--230.

\bibitem{Zagrovic:2003}
Bojan Zagrovic and Vijay Pande, \emph{Solvent viscosity dependence of the
  folding rate of a small protein: distributed computing study}, Journal of
  Computational Chemistry \textbf{24} (2003), 1432--1436.

\end{thebibliography}

\appendix

\section{Inclusion-Exclusion Volume and Surface Area Formulas}\label{inclusion_exclusion}

The equations in sections \ref{signed_volume} and \ref{equations_subsection} are 
extensions od the short inclusion-exclusion formulas for the atomic surface areas and 
volumes 
of a molecule discussed in \cite{Union},\cite{SFandV}, 
\cite{ShapeComp}, \cite{bio_solv}. We use the same notation as in \ref{signed_volume}.

\begin{figure}[h]
\begin{center}

\includegraphics[scale=.3]{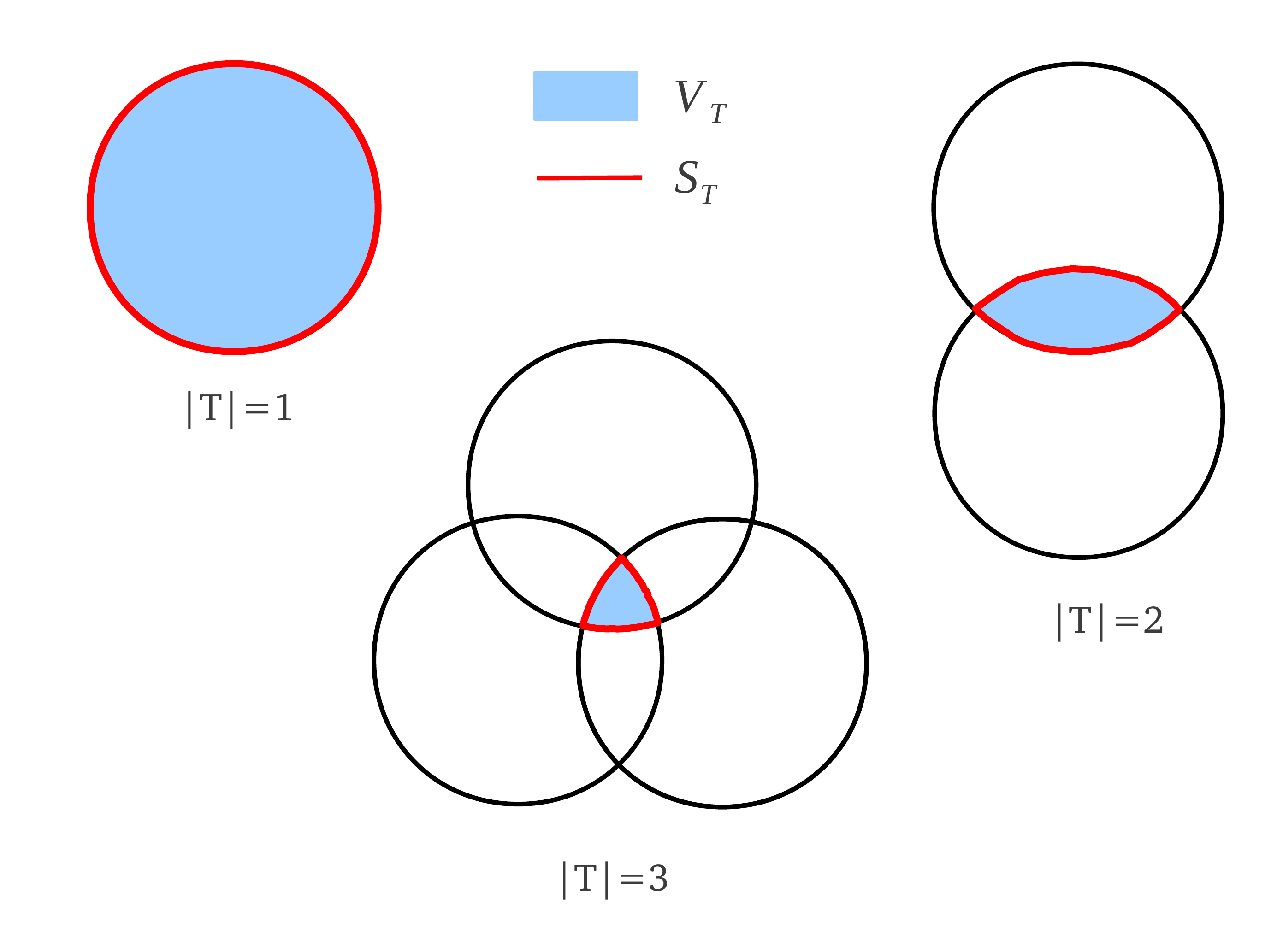}
\end{center}
\caption[Volume and surface area contributions from vertices, edges, and triangles.]{Volume (blue) and surface area (red) contributions 
from simplices of dimensions one, two, and three.}\label{V_T} 
\end{figure}

 The terms 
$S_T$ and $V_T$ are
the surface area and volume respectively of the intersection of the balls in $T$ (See Figure \ref{V_T}).
 The surface area and volume of the union of balls are \cite{Union},\cite{SFandV},   
\begin{eqnarray}\label{inclusion_exclusion_eq1}
\mathcal{S}&=&\sum_{\sigma_T \in \partial \mathcal{C}} (-1)^{k+1} c_T S_T, \quad \quad  |T|=k \\
\mathcal{V}&=&V+\sum_{\sigma_T \in \partial \mathcal{C}} (-1)^{k+1} c_T V_T, \quad \quad |T|=k 
\end{eqnarray}
where $V$ is the volume of the tetrahedra in the alpha complex and the coefficients, $c_T$, are given in Section \ref{signed_volume}.

\begin{figure}[h]
\begin{center}
\includegraphics[scale=.3]{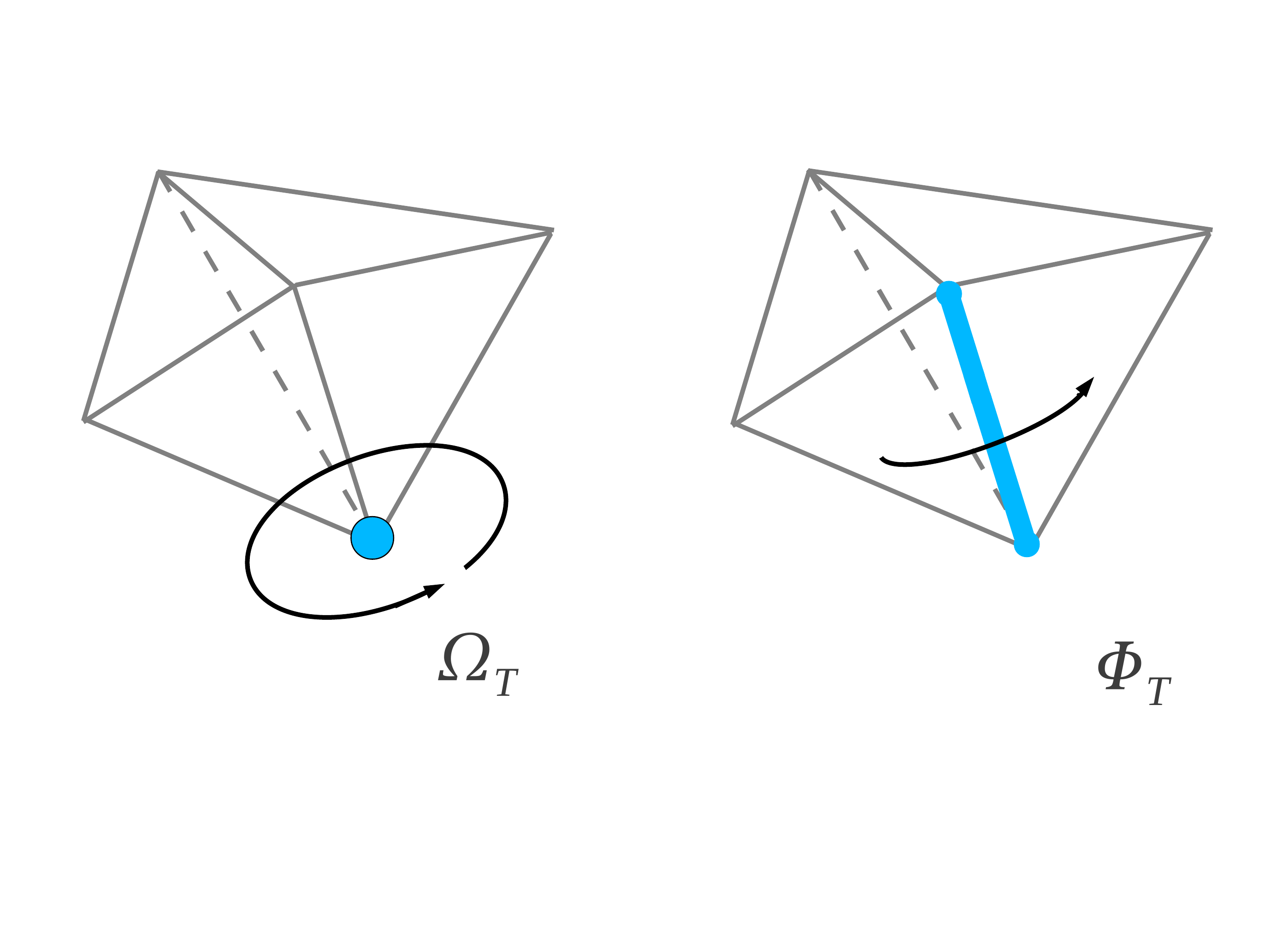}
\end{center}
\caption{Fractional outer solid angle and fractional outer dihedral angle.}
\label{c_T}
\end{figure}

The term $\SA$ gives the total surface area of the molecule, but we are also interested in the contribution of an individual atom, $p_i$, 
to the total surface area.  Call this 
term $\SA^i$.  Then 
\begin{equation}\label{surf_atom}
\SA^i=\sum_{\sigma_T \in \partial \mathcal{C}} (-1)^{k+1} c_T S^{(i)}_T, \quad \quad  |T|=k 
\end{equation}
where $S^{(i)}_T$ is the contribution of $S_T$ to $\SA^i$ with
\begin{equation}
 \sum_{i=1}^n\SA^i = \SA.
\end{equation}
These formulas will be given in subsequent sections. 
The sum \ref{surf_atom} may also be taken over all $\sigma_T \in \partial C$ such that $p_i \in T$ since $S^{(i)}_T=0$ if $p_i \notin T$. 

\subsection{Equations}

\subsubsection{$|T|=1$: Volume and Surface Area}

Consider $T=\{p_i\}$.  The formulas for the volume and surface area of a ball are 
\begin{eqnarray}
V_T&=&\frac{4}{3}\pi {p_i''}^{3/2} \\
S_T & = & 4\pi p_i'' \\
S^{(i)}_T & = & 4\pi p_i''.
\end{eqnarray}


Let $\mathcal{I}$ be the set of tetrahedra in $\mathcal{C}$ to which the vertex $p_i$ is incident. 
For $\sigma_I \in \mathcal{I}$ define $\omega_I$ as the normalized inner solid angle subtended by the 
tetrahedron $\sigma_I$ from the point $p'_i$.  Then 
\begin{eqnarray}
 \Omega_T& =& 1-\sum_{\sigma_{I} \in \mathcal{I}} \omega_T^I \\
\end{eqnarray}

\begin{figure}[h]
\begin{center}
 \includegraphics[scale=.25]{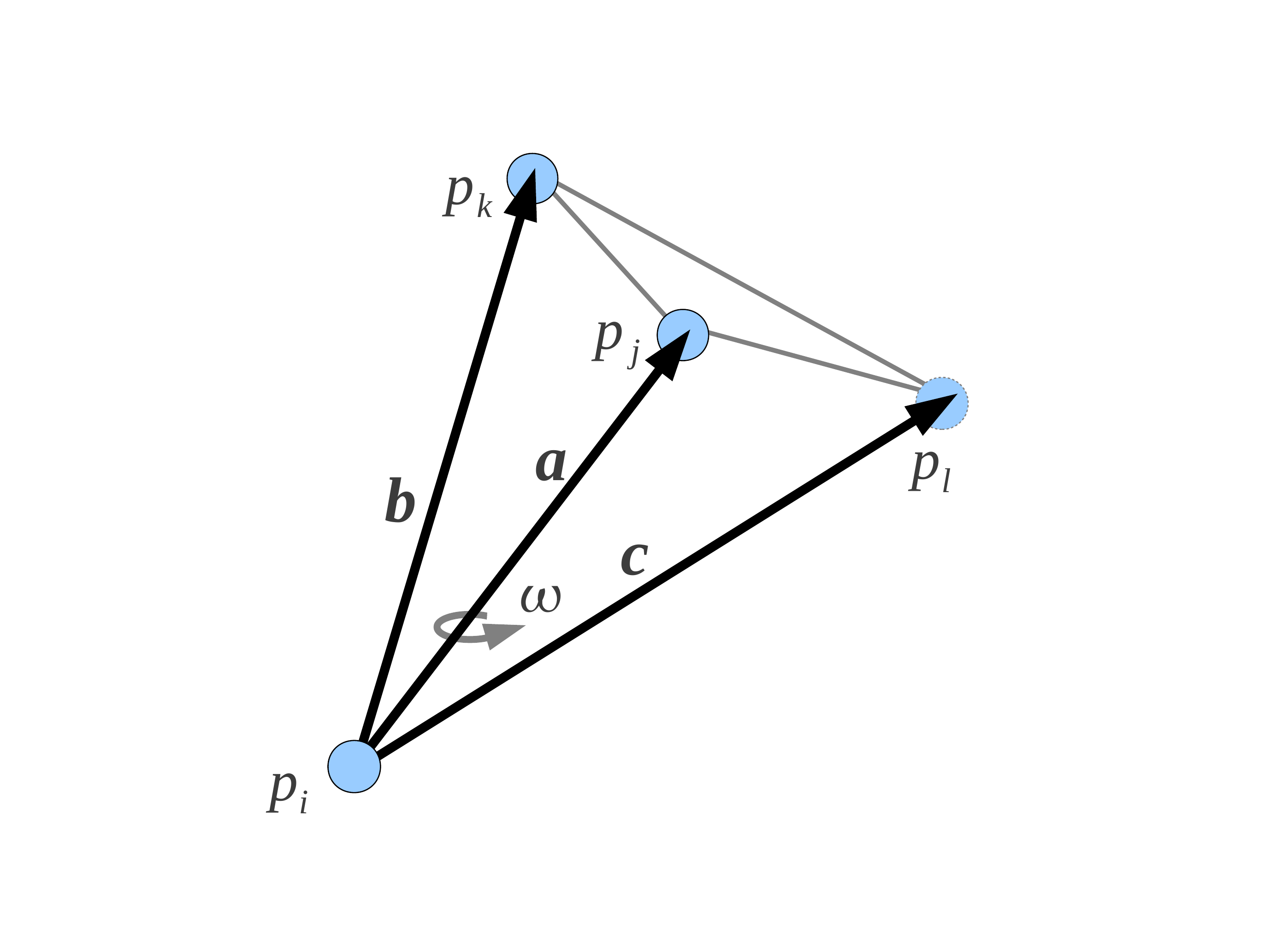}
\end{center}
\caption[Normalized inner solid angle.]{Normalized inner solid angle, $\omega$, subtended by
 $\mb{a}=p_j'-p_i'$, $\mb{b}=p_k'-p_i'$, and $\mb{c}=p_l'-p_i'$. }\label{pijkl}
\end{figure}

The normalized inner solid angle, $\omega$, of a tetrahedron subtended by the vectors 
$\mb{a}=p_j'-p_i'$, $\mb{b}=p_k'-p_i'$, and $\mb{c}=p_l'-p_i'$ (See Figure \ref{pijkl})
is given by the equation
\begin{equation}
  \omega  = \frac{1}{2\pi} \arctan\left( \frac{|\mathbf{a} \cdot 
(\mathbf{b} \times \mathbf{c})|}{ abc + (\mathbf a \cdot \mathbf b)c + 
(\mathbf a \cdot \mathbf c)b + (\mathbf b \cdot \mathbf c)a}\right)
\end{equation}
where $a=|\mathbf{a}|$ and likewise for $b$ and $c$ (See Figure \ref{pijkl}).

\subsubsection{$|T|=2$:  Volume and Surface Area}
Consider $T=\{p_i,p_j\}$.  The formulas for the volume and surface area of the intersection of the two balls, $p_i$ and $p_j$
are 
\begin{eqnarray} 
V_T& = &\pi h_i^2\bigg{(}\sqrt{p_i''}-\frac{h_i}{3}\bigg{)}+\pi h_j^2\bigg{(}\sqrt{p_j''}-\frac{h_j}{3}\bigg{)} \\
S_T^{(i)} & = & 2\pi\sqrt{p_i''} h_i \\
S_T^{(j)} & = & 2\pi\sqrt{p_j''} h_j \\
S_T & = & S_T^{(i)}+S_T^{(j)}
\end{eqnarray}
\begin{figure}[h]
\begin{center}
\includegraphics[scale=.25]{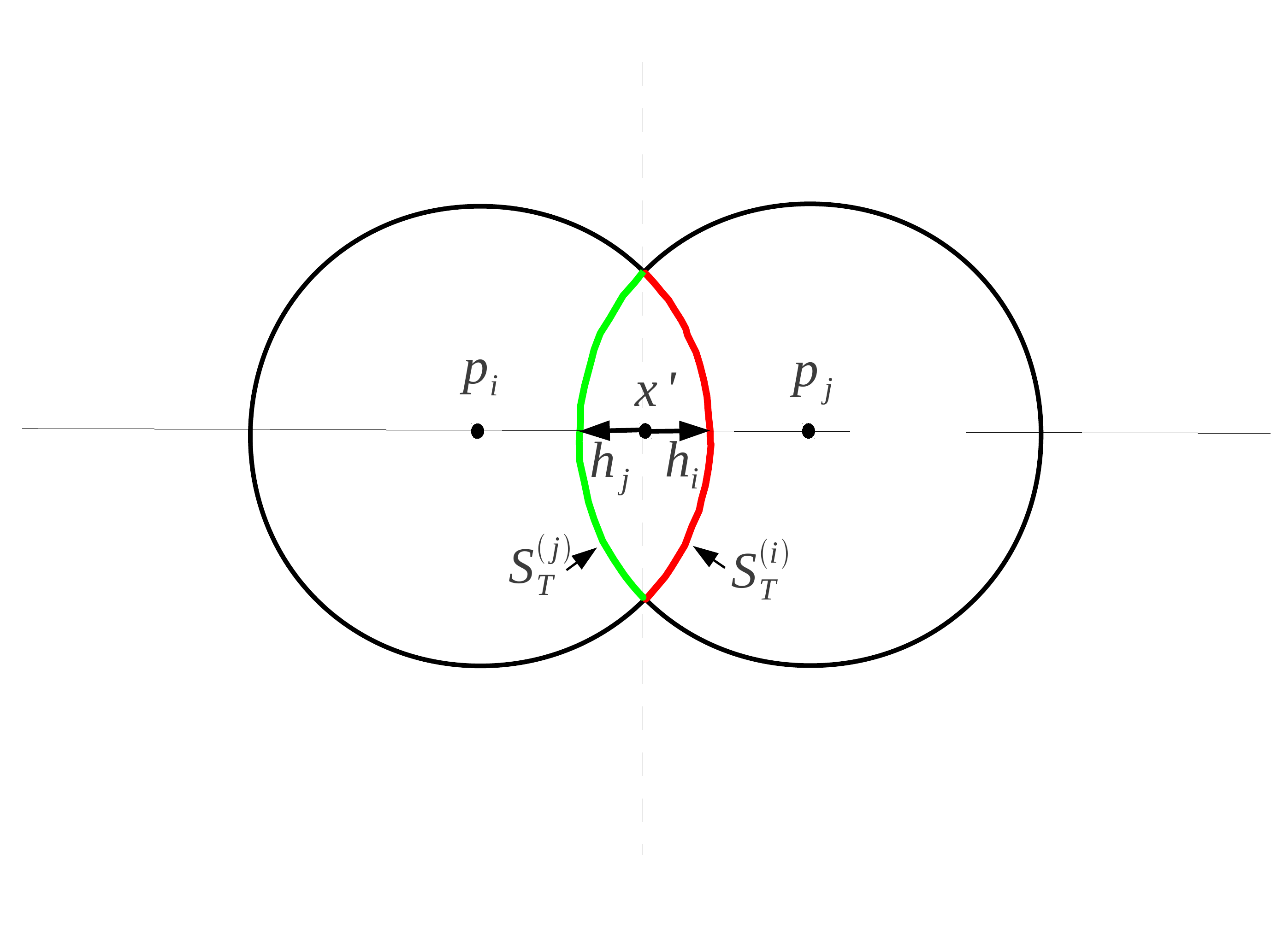}
\end{center}
\caption{Heights of the spherical caps and the partition of the surface
 area $S_T$ between two atoms.}\label{partition_SA}
\end{figure}
where $h_i$ and $h_j$ are the heights 
of the spherical caps of $p_i$ and $p_j$ (See Figure \ref{partition_SA}).  

The characteristic point, $\mb{x} \in \mathbb{R}^3 \times \mathbb{R}$,  
of the edge $\sigma_T$ satisfies
\begin{eqnarray}\label{VSA2}
 \Pi(\mb{p}_i,\mb{x})& = & 0 \nonumber \\
\Pi(\mb{p}_j,\mb{x}) & = & 0 \nonumber \\
 \mb{x'}&= &\mb{p}_i'+ t(\mb{p}_j'-\mb{p}_i') 
\end{eqnarray}
for some scalar $t$ (See Figure \ref{t_val}).  Equations \ref{VSA2} gives  
\begin{equation}
 p_i'^2-p_j'^2-2\mb{x'}\cdot(\mb{p}_j'-\mb{p}_i') -p_i''+p_j'' = 0 . \nonumber 
\end{equation}
Define $k=p_i''-p_j''$ and $\mb{a}=\mb{p}_j'-\mb{p}_i'$.  Then 
\begin{eqnarray}
t  &= &\frac{k-{p'_i}^2+{p'_j}^2-2\mb{p'}_i\cdot(\mb{p}_j'-\mb{p}_i')}{2a^2} \nonumber \\ 
& = & \frac{1}{2}\Bigg(\frac{k}{a^2}+1\Bigg)
\end{eqnarray}
and
\begin{eqnarray}
 h_i & =& \sqrt{p_i''}-sgn(t)|\mb{x'}-\mb{p'}_i|  \\ \nonumber
 h_j &= &\sqrt{p_j''}-sgn(1-t)|\mb{x'}-\mb{p'}_j|. \nonumber
\end{eqnarray}


Let $\mathcal{I}$ be the set of tetrahedra in $\mathcal{C}$ to which the edge $\sigma_T$
is incident.  For $\sigma_I \in \mathcal{I}$ define $\phi_I$ as the normalized inner dihedral angle 
of $\sigma_I$ along $\sigma_{T}$.  Then
\begin{equation}
 \Phi_T = 1 - \sum_{\sigma_I \in \mathcal{I}} \phi_T^I
\end{equation}
The normalized dihedral angle between planes with normals $\mb{n}_k$ and $\mb{n}_l$ is
\begin{equation}
 \phi=\frac{\arccos (\mb{n}_k \cdot \mb{n}_l)}{2\pi}.
\end{equation}

\begin{figure}[h]
\begin{center}
\includegraphics[scale=.2]{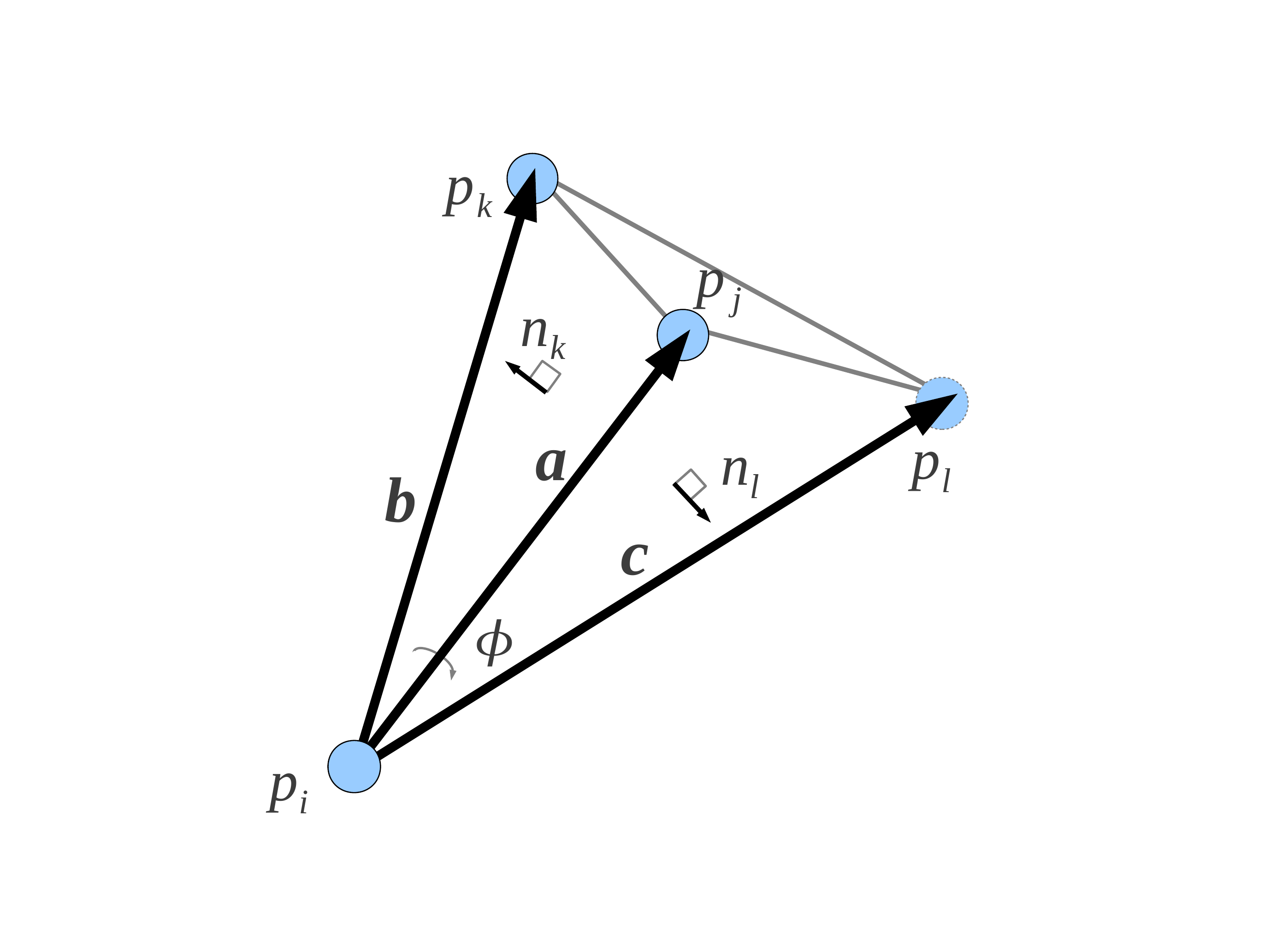}
\end{center}
\caption{Normalized dihedral angle $\phi$ between planes with normals 
$\mb{n}_k$ and $\mb{n}_l$.}\label{phi_normal}
\end{figure}

Assume plane $k$ is defined by the vectors $\mb{a}=\mb{p'}_j-\mb{p'}_i$ and 
$\mb{b}=\mb{p'}_k-\mb{p'}_i$ and the plane $l$ is defined by the vectors
$\mb{a}$ and $\mb{c}=\mb{p'}_l-\mb{p'}_i$.  Then
\begin{eqnarray}
 \mb{n}_k& =& \frac{\mb{a} \times \mb{b}}{|\mb{a}\times \mb{b}|} \nonumber \\
\mb{n}_l & = & \frac{\mb{a} \times \mb{c}}{|\mb{a}\times \mb{c}|}.
\end{eqnarray}

\subsubsection{$|T|=3$: Volume and Surface Area}\label{tri_eqns}

Consider $T = \{p_i, p_j, p_k\}$.
The volume and surface area of the common intersection of three balls can be written as a 
weighted sum of the surface area of the single and the double intersections.   
If $p_i$, $p_j$, and $p_k$ have a non-empty intersection then there
are two points in common with the surfaces of all three balls.  Call one of these points $\mb{x}'$, define 
$p_x=(\mb{x}',0) \in \mathbb{R}^3\times \mathbb{R}$,
and let $T_x= \{p_i, p_j, p_k,p_x\}$.  Let $S_2$ be the set of edges defined by $\sigma_{T_x}$ and
$S_1$ the set of vertices in $\sigma_{T_x}$.

\begin{figure}[h]
\begin{center}
 \includegraphics[scale=.3]{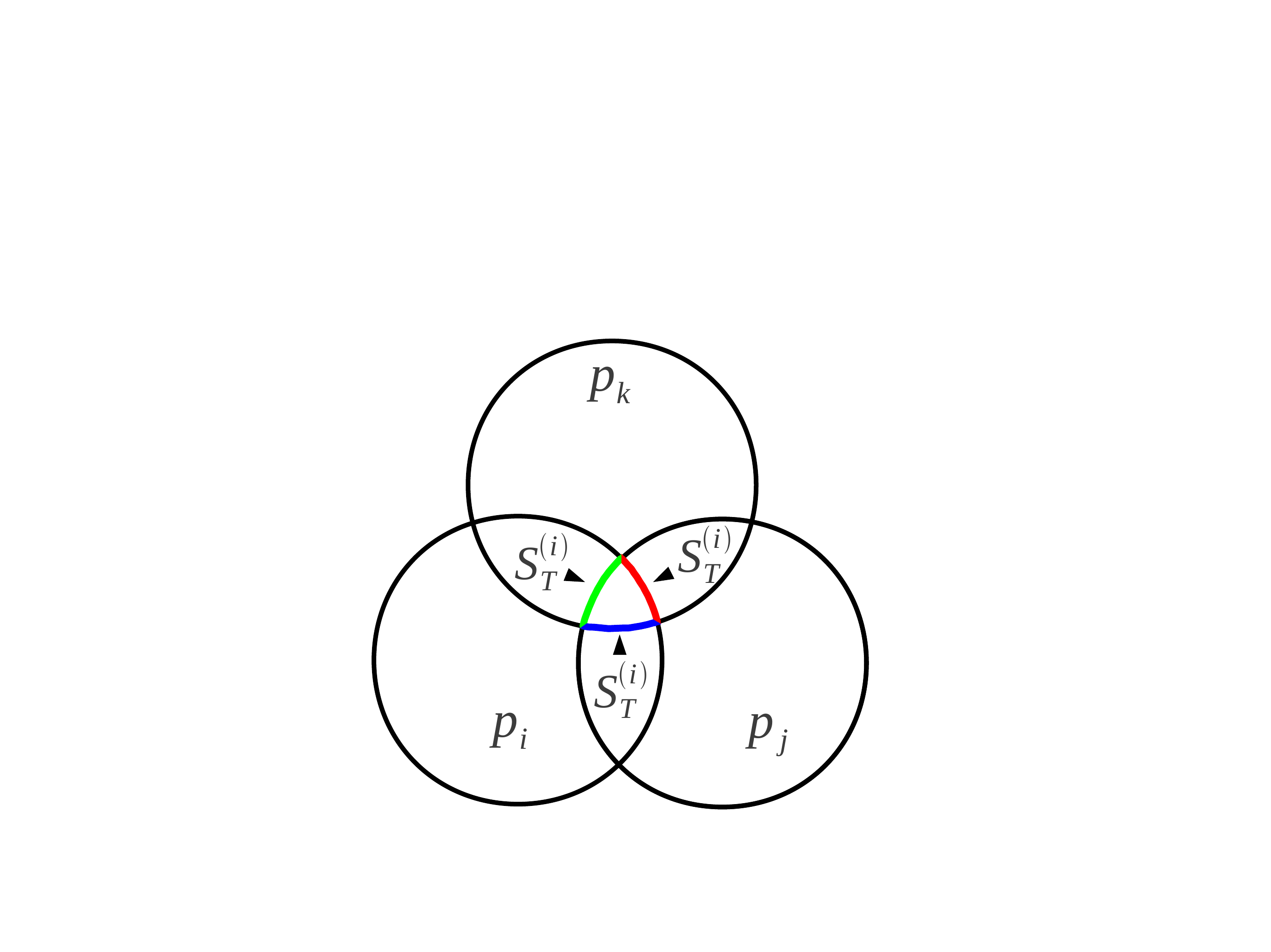}
\end{center}
\caption{Partition of $S_T$ between the three atoms.}\label{partition_atoms3}
\end{figure}

The volume and surface areas (See Figure \ref{partition_atoms3}) of the intersection of $p_i$, $p_j$,
 and $p_k$ are given by \cite{Vol3}
\begin{eqnarray}
\frac{1}{2}V & = & V_{T_c}+\sum_{\sigma_t \in S_2} \Phi_{t}V_{t}-\sum_{\sigma_t \in S_1} \Omega_t V_t \quad k,l = (1,2,3) \\
  \frac{1}{2}\SA^{(i)}_T & = &\Phi_{\{i,j\}}S^{(i)}_{\{i,j\}}+\Phi_{\{i,k\}}S^{(i)}_{\{i,k\}}-\Omega_{\{i\}} S_{\{i\}}\\
 \frac{1}{2}\SA^{(j)}_T & = &\Phi_{\{j,k\}}S^{(j)}_{\{j,k\}}+\Phi_{\{j,i\}}S^{(j)}_{\{j,i\}}-\Omega_{\{j\}} S_{\{j\}}\\
\frac{1}{2} \SA^{(k)}_T & = &\Phi_{\{k,i\}}S^{(k)}_{\{k,i\}}+\Phi_{\{k,j\}}S^{(k)}_{\{k,j\}}-\Omega_{\{k\}} S_{\{k\}}\\
\frac{1}{2}\SA_T & = & \SA^{(i)}_T+\SA^{(j)}_T+\SA^{(k)}_T
\end{eqnarray}
where $\Phi_{\{i,j\}}$ is the normalized dihedral angle of $\sigma_{T_C}$ along the edge $\sigma_{\{i,j\}}$, 
$\Omega_i$ is the normalized solid angle of $\sigma_{T_c}$ subtended from $p'_i$, and similarly for other combinations 
$i$, $j$, and $k$.


The point $\mb{x}$ satisfies the following equations
\begin{eqnarray}
 |\mb{p'}_i-\mb{x}'|^2-p_i'' & = & 0 \label{e1}   \\ 
  |\mb{p'}_j-\mb{x}'|^2-p_j'' & = & 0 \label{e2}  \\
 |\mb{p'}_k-\mb{x}'|^2-p_k'' & = & 0 \label{e3} 
\end{eqnarray}
Let $\mb{a} = \mb{p_j}'-\mb{p}_i'$ and $\mb{b} = \mb{p}_k'-\mb{p}_i'$, which gives the normal 
to the plane that contains $\sigma_T$ as $\mb{n} =  \mb{a} \times \mb{b}$.  
The characteristic point, $\mb{x}_c$, of the triangle $\sigma_T$ was found in the computation of 
the alpha complex and satisfies $\mb{x} = \mb{x}_c + h \mb{n}$ where $h$ is the height of $\mb{x}$ above the plane containing $\sigma_T$.
Plugging this into equation \ref{e1} gives 

\begin{equation}
|\mb{p}_i'-\mb{x}_c|^2-2h(\mb{p}_i'-\mb{x}_c)\cdot \mb{n}+h^2 n^2-p_i'' =0  
\end{equation}
The vector $(\mb{p}_i'-\mb{x}_c)$ is orthogonal to $\mb{n}$ which gives
\begin{equation}\label{h_eqn}
 h = \frac{\sqrt{p_i''-|\mb{p}_i'-\mb{x}_c|^2}}{n} = \frac{\sqrt{-\alpha_{\sigma_T}}}{n}
\end{equation}
where $\alpha_{\sigma_T}$ is the size of the triangle $\sigma_T$.

Given a triangle, $\sigma_T$, its coefficient is given by 
\begin{equation}
 c_T = \begin{cases}
        1, & \text{if } \sigma_T \text{ is singular} \\
        \frac{1}{2} & \text{if } \sigma_T \text{ is regular}
       \end{cases}
\end{equation}
The triangle, $\sigma_T$, transitions from singular to regular when an incident tetrahedron 
becomes part of the alpha complex.  

\subsection{$|T|=4$: Volume}
Consider $T=\{p_i,p_j,p_k,p_l\}$.  Let $\mb{a}=p_j'-p_i'$, $\mb{b}=p_k'-p_i'$, and $\mb{c}=p_l'-p_i'$.
The volume of the tetrahedron $\sigma_T$ is given by
\begin{equation}
 V_{tetra} = \frac { |\mathbf{a} \cdot (\mathbf{b} \times \mathbf{c})| } {6}.
\end{equation}

\section{Exterior Facet Area Computation Pseudocode}
\verbatiminput{Laguerre_stuff/pseudocode_3.f90}

\section{Glossary}

\begin{description}


\item[$\pi$] \hfill \\ 
Power distance between weighted point and unweighted point; $\pi(p,x')=|p'-x'|^2-p''$.


\item[$\sigma_T$] \hfill \\ 
A simplex which is the convex hull of point centers in set $T$.

%

\item[A]

 \item[$\mathcal{A}$ ] \hfill \\  
 Set of weighted data points (spheres) which represents atoms in a molecule.
\item[$\mathcal{A}'$] \hfill \\ 
Set of data point centers; $\mathcal{A}'= \{ p' \text{  such that } p \in \mathcal{A} \}$.
\item[$\mathcal{A}(w)$] \hfill \\ 
Set of expanded points, $\mathcal{A}(w)=\{ p(w) \text{ such that } p \in \mathcal{A} \}$.

\item[B]
\item[$\mathcal{B}(w)$] \hfill \\  
Space filling model of $\mathcal{A}(w)$.  Equal to $\bigcup_{p_i \in \mathcal{A}} B_i(w)$.

\item[$B_i(w)$] \hfill \\ 
Closed ball with center at $p_i'$ and weight $p_i''$.

\item[C]
\item[$\mathcal{C}(w)$]  \hfill \\ 
Alpha complex of data set with $\alpha=w$.

\item[$\partial \mathcal{C}$] \hfill \\ 
Boundary of the alpha complex $\mathcal{C}$.


\item[E] 
\item[$e_{ij}$] \hfill \\
Edge connecting the centers of $p_i$ and $p_j$.

\item[F]
\item[$F_T$] \hfill \\ 
Surface area of the intersection of the Laguerre facet corresponding to simplex $\sigma_T$ 
with the interior of the alpha complex.
 
%
%
%
%

%

\item[L]

\item[$\mathcal{L}(\mathcal{A})$] \hfill \\	
Laguerre diagram of the data set $\mathcal{A}$.

%

\item[$L_i$] \hfill \\ 
Laguerre cell of atom $i$. 

\item[$L_i(w)$] \hfill \\ 
Laguerre cell of $p_i(w)$ in $\mathcal{A}(w)$.

\item[$L_{ij}$] \hfill \\ 
Laguerre facet corresponding to edge $e_{ij}$.


\item[$LI(w)$] \hfill \\ 
Set of Laguerre-Intersection cells.

\item[$LI_i(w)$] \hfill \\ 
Laguerre-Intersection cell of atom $i$ with weight $w$.

\item[$LIS_i(w)$] \hfill \\ 
Surface area of $LI_i(w)$.

\item[$LIS_T^{(i)}$] \hfill \\ 
Contribution of the simplex $\sigma_T$ to $LIS_i$.

\item[$LIV_i(w)$] \hfill \\ 
Volume of $LI_i(w)$.

\item[$LIV_T^{(i)}$] \hfill \\ 
Contribution of the simplex $\sigma_T$ to $LIV_i$.

\item[$LIV_i^{(i)}$] \hfill \\ 
Equivalent to $LIV_T^{(i)}$ for $T=\{p_i\}$.

\item[$LIV_{ij}^{(i)}$] \hfill \\ 
Equivalent to $LIV_T^{(i)}$ for $T=\{p_i,p_j\}$.

\item[$LIV_{ijk}^{(i)}$] \hfill \\ 
Equivalent to $LIV_T^{(i)}$ for $T=\{p_i,p_j,p_k\}$.

\item[$LS_i$] \hfill \\ 
Surface area of the Laguerre cell of atom $i$.

\item[$LS_{ij}$] \hfill \\ 
Surface area of facet $L_{ij}$.

\item[$LV_i$] \hfill \\ 
Volume of the Laguerre cell of atom $i$.

\item[$LV_{ij}^i$] \hfill \\ 
Volume contribution of the edge $e_{ij}$ to the Laguerre cell of atom $i$.

%

\item[P]


\item[$p=(p',p'')$ ] \hfill \\ 
Weighted point (sphere) in $\mathbb{R}^3 \times \mathbb{R}$ with center $p'$ and weight (radius squared) $p''$.  
\item[$p'$ ] \hfill \\  
Unweighted point or location in $\mathbb{R}^3$. 
\item[$p''$ ] \hfill \\  
Weight or radius squared of point $p$. This can also be written as $w$. 

\item[$p(w)$] \hfill \\ 
Expanded point $p(w)=(p',p''+w)$

\item[$P_T^{(i)}$] \hfill \\ 
Pyramidal volume contribution of $\sigma_T$ to $LIV_i$.

\item[S]


\item[$S_i(w)$] \hfill \\ 
Accessible surface area of atom $i$ with solvent weight $w$.

\item[$S_T$] \hfill \\ 
Surface area of the intersection of balls represented by points in $T$. 

\item[$S_i^{(i)}$] \hfill \\ 
Equivalent to $S_T^{(i)}$ where $T=\{p_i\}$.

\item[$S_T^{(i)}$] \hfill \\ 
The contribution of $S_T$ to $S_i$.

\item[T]

\item[$\mathcal{T}(\mathcal{A})$] \hfill \\ 
Weighted Delaunay (Regular) tetrahedrization of the data set $\mathcal{A}$.
\item[$T$] \hfill \\ 
Set of points corresponding to simplex in $\mathcal{A}$.   
\item[$|T|$] \hfill \\  
Number of elements in the set $T$.

\item{$t_{ijk}$} \hfill \\ 
A triangle with vertices at $p_i'$, $p_j'$, $p_k'$. 

\item[V]

\item[$v_i$] \hfill \\
Vertex.

\item[$V_T$] \hfill \\
Volume of the intersection of balls represented by points in $T$.

\item[W]

\item[w] \hfill \\
Weight or radius squared of a data point


\item[X]

%

\item[$x_{ij}$] \hfill \\ 
Characteristic point of the edge $e_{ij}$.

\item[$x_{ijk}$] \hfill \\ 
Characteristic point of the triangle $t_{ijk}$.

\item[$x_T$] \hfill \\ 
Characteristic point of the simplex $\sigma_T$.  $x_T=(x_T',x_T'')$ where $x_T'$ is the center of the characteristic 
point and the weight $x_T''$ is the size of the simplex.

\end{description}


\end{document}